%% file: main.tex
\documentclass[conference]{IEEEtran}
\IEEEoverridecommandlockouts
\usepackage{cite}
\usepackage{amsmath,amssymb,amsfonts}
\usepackage{algorithmic}
\usepackage{algorithm}
\usepackage{graphicx}
\usepackage{textcomp}
\usepackage{booktabs}
\usepackage{multirow}
\usepackage{xcolor}
\usepackage{subcaption}
\usepackage{bm}
\usepackage[
            colorlinks,
            linkcolor=red, 
            anchorcolor=green,
            citecolor=blue,
            ]{hyperref}

\def\BibTeX{{\rm B\kern-.05em{\sc i\kern-.025em b}\kern-.08em
    T\kern-.1667em\lower.7ex\hbox{E}\kern-.125emX}}

\newcommand{\ie}{\emph{i.e., }}
\newcommand{\eg}{\emph{e.g., }}

\newcommand{\wrt}{\emph{w.r.t. }}

\newcommand{\Mat}[1]{\mathbf{#1}}

\begin{document}
\renewcommand{\arraystretch}{1.2}

\title{DiffsFormer: A Diffusion Transformer on Stock Factor Augmentation
\thanks{$^\dagger$ Work done at Alibaba Group.}
}


\author{\IEEEauthorblockN{Yuan Gao\IEEEauthorrefmark{4}$^\dagger$, Haokun Chen\IEEEauthorrefmark{3}, Xiang Wang\IEEEauthorrefmark{4}, Zhicai Wang\IEEEauthorrefmark{4}, \\
Xue Wang\IEEEauthorrefmark{3}, Jinyang Gao\IEEEauthorrefmark{3}, Bolin Ding\IEEEauthorrefmark{3}}
\IEEEauthorblockA{\IEEEauthorrefmark{4}University of Science and Technology of China, Hefei, China\\
yuanga@mail.ustc.edu.cn, xiangwang1223@gmail.com, wangzhic@mail.ustc.edu.cn}
\IEEEauthorblockA{\IEEEauthorrefmark{3}DAMO Academy, Alibaba Group, Hangzhou, China\\
\{hankel.chk,xue.w,jinyang.gjy,bolin.ding\}@alibaba-inc.com}
}


\maketitle

\input{chapters/0_abstract}
\input{chapters/1_intro}
\input{chapters/2_related}
\input{chapters/3_background}
\input{chapters/4_method}
\input{chapters/5_experiment}
\input{chapters/6_conclusion}

\bibliographystyle{IEEEtran}
\bibliography{reference}

\end{document}

%% file: chapters/0_abstract.tex
\begin{abstract}
Machine learning models have demonstrated remarkable efficacy and efficiency in a wide range of stock forecasting tasks. However, the inherent challenges of data scarcity, including low signal-to-noise ratio (SNR) and data homogeneity, pose significant obstacles to accurate forecasting. 
To address this issue, we propose a novel approach that utilizes artificial intelligence-generated samples (AIGS) to enhance the training procedures.
In our work, we introduce the \underline{Diff}usion Model to generate \underline{s}tock factors with Trans\underline{former} architecture (DiffsFormer). DiffsFormer is initially trained on a large-scale source domain, incorporating conditional guidance so as to capture global joint distribution. When presented with a specific downstream task, we employ DiffsFormer to augment the training procedure by editing existing samples. This editing step allows us to control the strength of the editing process, determining the extent to which the generated data deviates from the target domain.
To evaluate the effectiveness of DiffsFormer augmented training, we conduct experiments on the CSI300 and CSI800 datasets, employing eight commonly used machine learning models. The proposed method achieves relative improvements of 7.2\% and 27.8\% in annualized return ratio for the respective datasets. Furthermore, we perform extensive experiments to gain insights into the functionality of DiffsFormer and its constituent components, elucidating how they address the challenges of data scarcity and enhance the overall model performance.

Our research demonstrates the efficacy of leveraging AIGS and the DiffsFormer architecture to mitigate data scarcity in stock forecasting tasks. The ability to generate realistic stock factors coupled with the control offered by the editing step presents a promising avenue for improving forecasting accuracy. The experimental results validate the potential of our proposed approach and contribute to a deeper understanding of its underlying mechanisms.

\end{abstract}


%% file: chapters/1_intro.tex
\section{Introduction}

Accurate stock forecasting plays a crucial role in effective asset management and investment strategies \cite{zou2022stock}. Its objective is to predict future stock behavior (\eg return ratios or prices) by analyzing relevant historical factors. Previous research has explored various machine learning techniques, such as SFM \cite{DBLP:conf/kdd/ZhangAQ17}, ALSTM \cite{DBLP:conf/ijcai/FengC0DSC19}, and HIST \cite{DBLP:journals/corr/abs-2110-13716}. However, achieving desirable performance with these methods often requires an ample supply of high-quality data. The challenges posed by high random and homogeneous data make it difficult to meet the requirements for data quality, resulting in elevated forecasting errors and increased uncertainty.


\begin{figure}[t]
    \centering
    \begin{subfigure}[b]{\columnwidth}
        \centering
        \includegraphics[width=\textwidth]{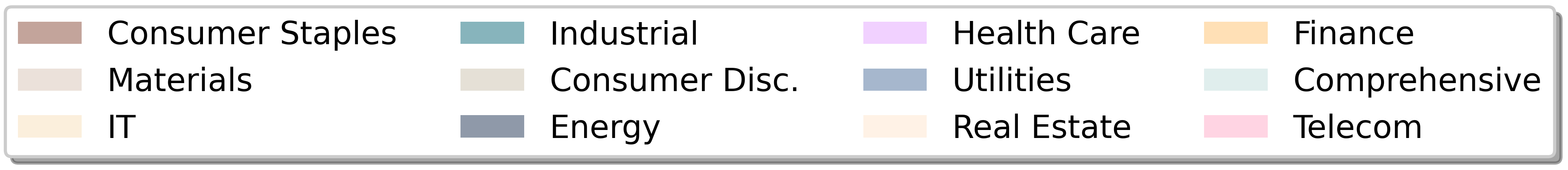}
    \end{subfigure}

    \smallskip
        
    \begin{subfigure}[b]{0.5\columnwidth}
        \centering
        \includegraphics[width=\textwidth]{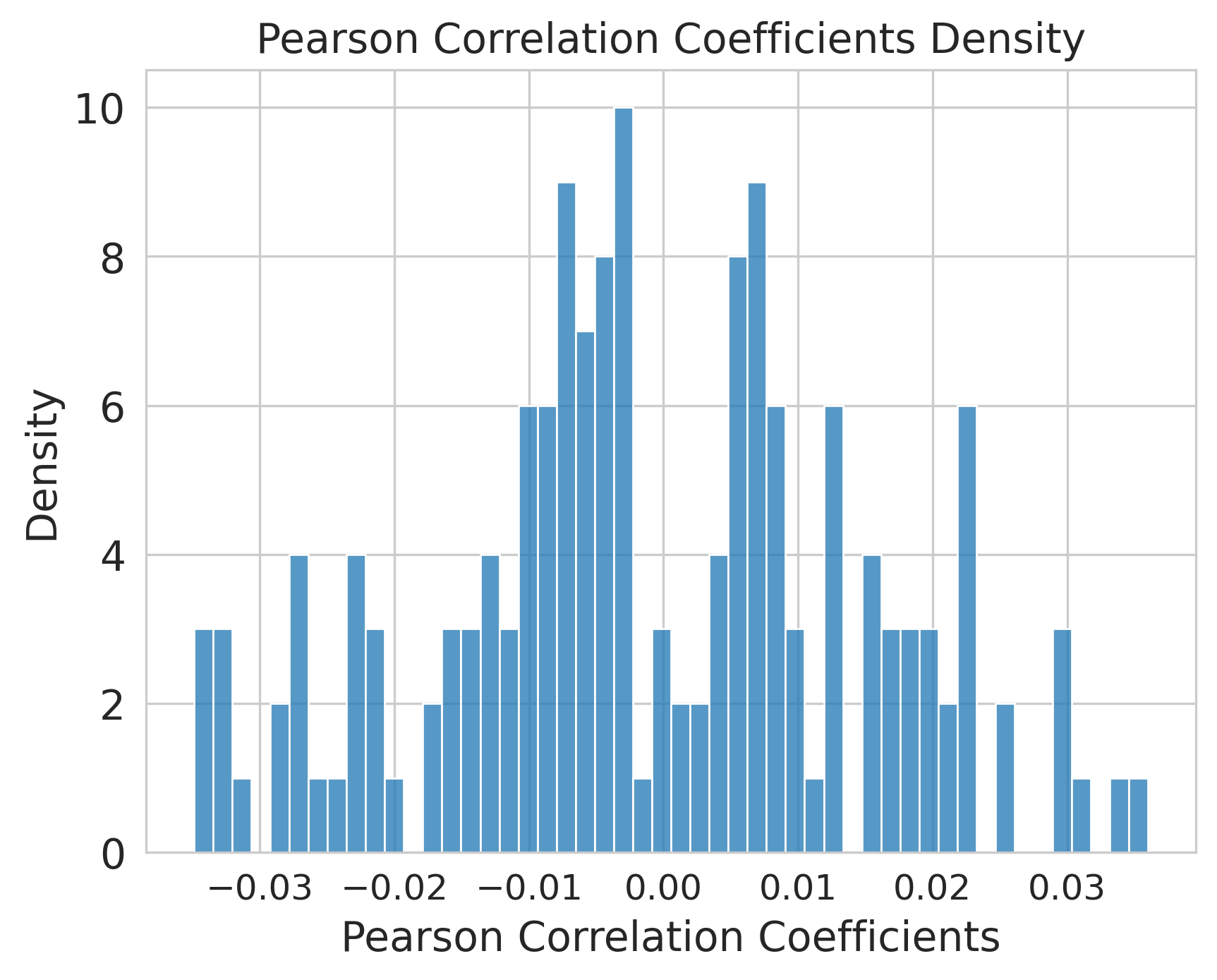}
        \caption{Data SNR}
        \label{fig:snr}
    \end{subfigure}%
    \begin{subfigure}[b]{0.5\columnwidth}
        \centering
        \includegraphics[width=\textwidth]{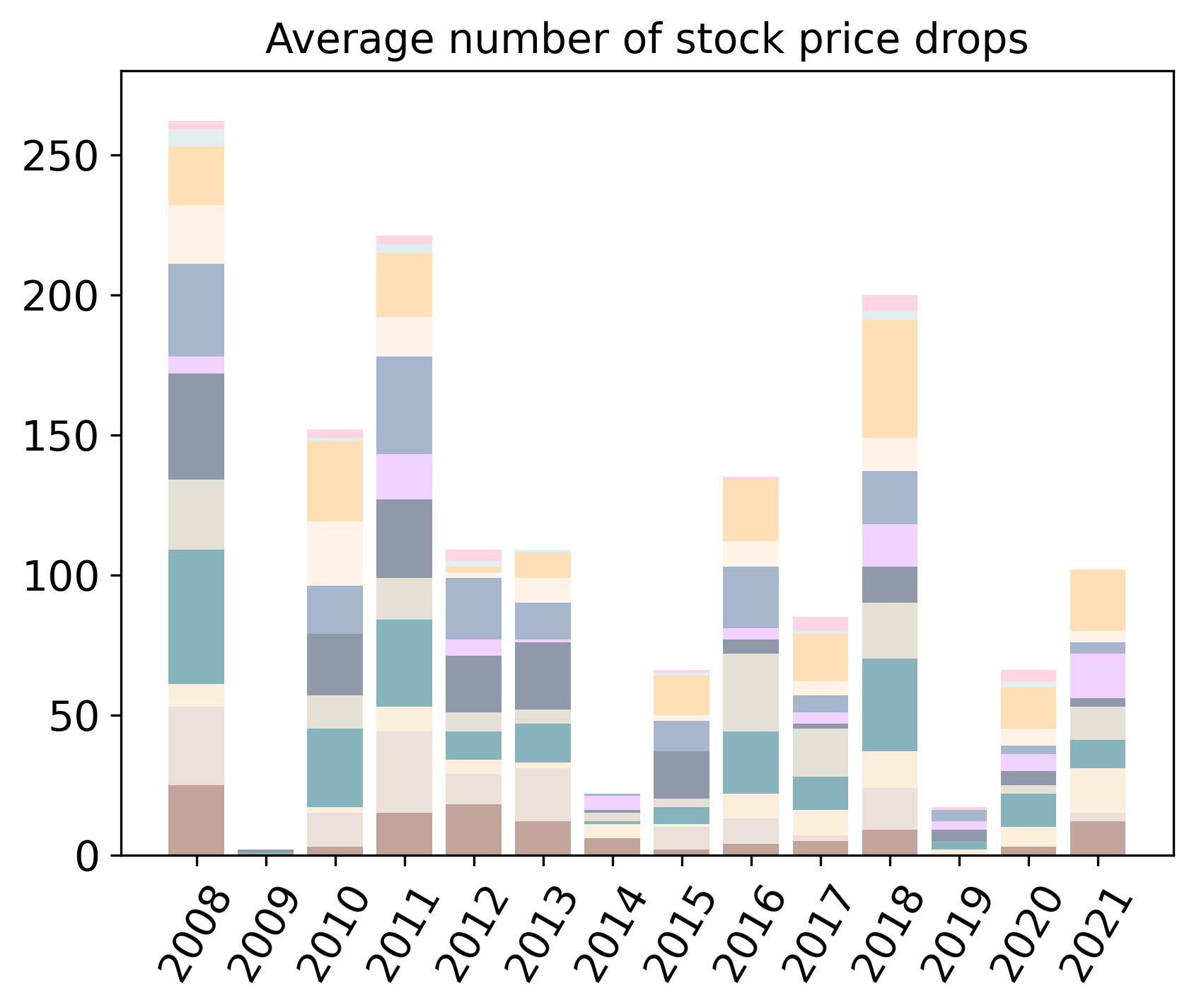}
        \caption{Data Homogeneity}
        \label{fig:homogeneity}
    \end{subfigure}

    \caption{(a) Pearson Correlation Coefficients between return ratio and stock factors are low. (b) Average number of large price drop stocks split by sectors. Stocks within the same industry sector tend to perform similarly.}
    \label{fig:data-scarcity}
    \vspace{-15pt}
\end{figure}

Data scarcity in stock forecasting tasks is commonly characterized from two aspects: \textit{signal-to-noise ratio (SNR)} and \textit{data homogeneity}. (1) We first delve into the relationship between stock factors and the return ratio, providing insights into SNR. Figure \ref{fig:snr} visualizes the Pearson Correlation Coefficients between stock factors and the return ratio, revealing a weak correlation (absolute value less than 0.03) indicative of a low SNR of the factors. This weak correlation is often attributed to randomness and non-stationary speculative behaviors. (2) We explore the behavior of stocks within industry sectors to show data homogeneity's implications. We discover that stocks within the same industry sector exhibit similar behavior, as shown in Figure \ref{fig:homogeneity}, which reports the average number of large price drop stocks categorized by sectors. As a result, this homogeneity leads to a reduction in the availability of stocks with distinctive informational characteristics. Data scarcity presents inherent challenges that can lead to overfitting, where models exhibit an increased risk of learning shortcuts and spurious correlations, ultimately impacting their predictive performance. The limited availability of data poses a significant hurdle in achieving effective generalization between the training and testing datasets, thereby resulting in compromised overall performance.
One straightforward solution to overcome data scarcity is factor augmentation, a technique that involves augmenting the training dataset by introducing minor changes or generating new data points based on specific factors. Drawing inspiration from successful applications of Diffusion Models (DMs) in various domains, such as text-to-image conversion \cite{nichol2021glide, ramesh2022hierarchical}, time-series imputation \cite{tashiro2021csdi}, and waveform generation \cite{chen2020wavegrad}, we explore the utilization of DMs for stock forecasting. DMs are generative models consisting of two stages: a diffusion process and a denoising process. The diffusion process parameterizes a Markov chain that progressively introduces noise to the factors until reaching a state of pure noise \cite{ho2020denoising}. Subsequently, during the denoising process, the model aims to restore the original data by predicting the noise generated through the diffusion process. In this study, we look back 8 days and organize recent stock factors as a sequence, leveraging DMs based on transformer architectures \cite{peebles2022scalable, tashiro2021csdi} to do factor augmentation. We expect that by incorporating augmented factors, our proposed model will exhibit enhanced resilience to data scarcity in the field of stock forecasting.

\begin{figure}[t]
\centerline{\includegraphics[width=\columnwidth]{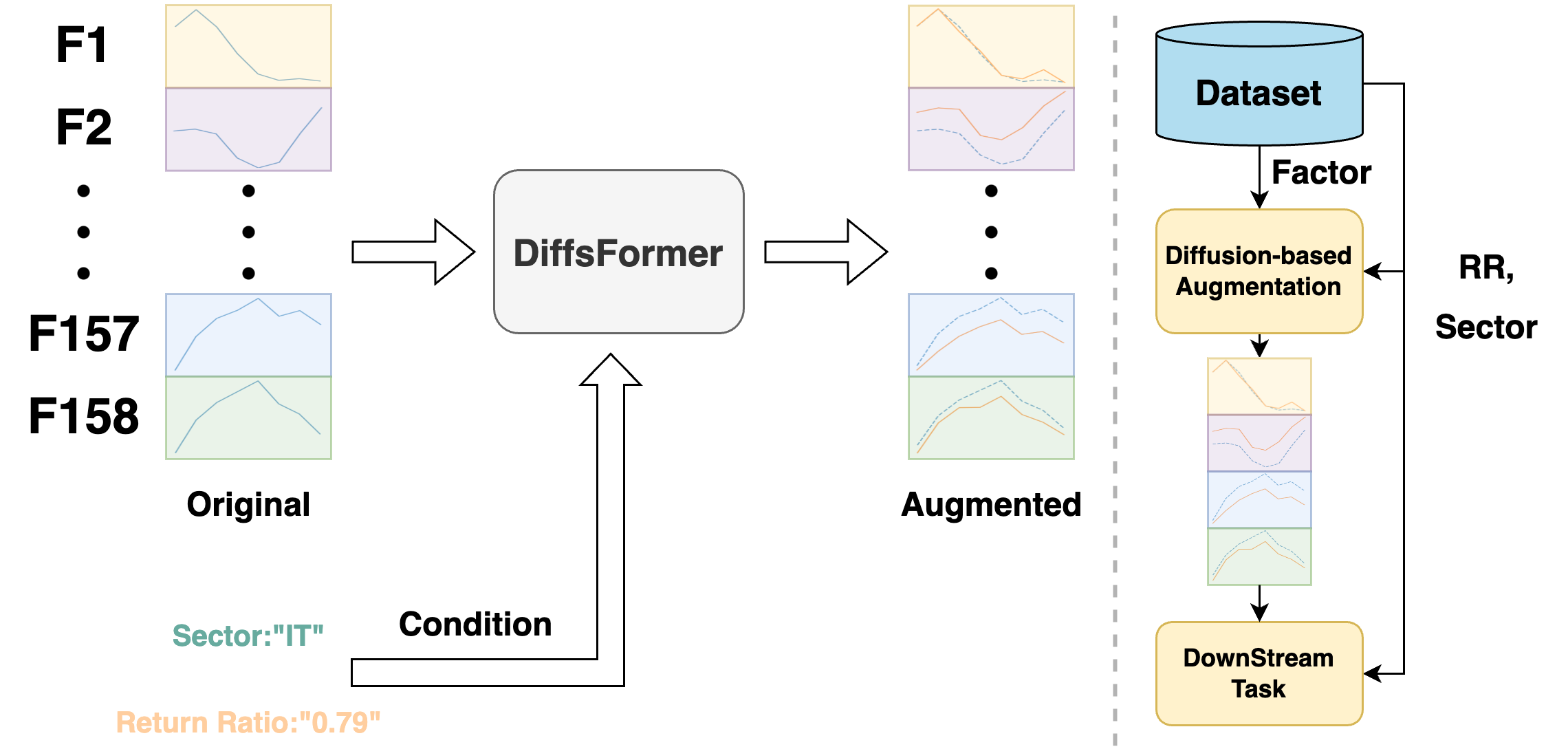}}
\caption{An illustration of DiffsFormer. F refers to ``factors'', such as the open, close, lowest, and highest prices of a stock.}
\vspace{-10pt}
\label{fig:teaser}
\end{figure}

Incorporating DMs for stock factor augmentation presents non-trivial challenges, particularly in the assignment of corresponding labels to the generated factors. While one possible approach is to directly treat labels as a stock factor dimension, this method carries the risk of generating inaccurate results. Explicitly generating labels presents challenges due to the difficulty of accurately matching factors with their corresponding labels, particularly considering the unpredictability of return ratios. To overcome this challenge, we propose to adapt DMs from generation task to supervised-learning task by utilization of a flexible predictor-free conditional factor generator \cite{ho2022classifier, nichol2021glide, peebles2022scalable}. The generator is trained with labels as the conditioning, allowing us to anticipate that the generated factors will share the same labels as the original factors. Additionally, we explore the utilization of other types of conditionings (\ie sectors) to further enhance accuracy. An illustration is shown in Figure \ref{fig:teaser}.


To distill new knowledge and information (\ie stocks in other markets) during augmentation, our proposed framework incorporates transfer learning. Initially, the DM is trained in a large source domain with a diffusion step denoted as 
$T$; during the inference process, instead of generating from a standard normal distribution, we start from data points in the target domain, corrupt them and then denoise to obtain a new data point that is likely in the target domain. However, due to the requirement of a large $T$ for DMs to reach a state of pure noise, the low SNR prevalent in stock data poses challenges in achieving accurate recovery as $T$ increases. In practical implementation, we limit the corruption of data to a smaller number of steps denoted as $T' \ll T$, which we refer to as the editing step. In addition, we have discovered a significant benefit resulting from this mechanism. The phenomenon known as data collision occurs when multiple models utilize the same copy of data simultaneously. In such cases, a surge of capital flowing into the market can lead to the failure of these models in achieving their investment objectives \cite{jacobs2020anomalies}. However, through the editing of factors, this issue can be effectively mitigated by introducing new copies of data. To summarize, our approach involves editing samples in the target domain using knowledge derived from a larger source domain. 

We also propose several improvements to increase the model's efficiency and decrease the volatility. It is obvious that there is no need to optimize the DM for $t > T'$. On top of that, we initialize the DM with $T$ to ensure its correctness, but sample training steps $t$ from ${\rm Uniform} \{1, 2, \cdots, T'\}$ instead of ${\rm Uniform} \{1, 2, \cdots, T\}$. Additionally, we leverage the training loss as a proxy and introduce stronger noise to data points with lower training loss. This loss-guided noise addition mechanism aims help decrease model volatility by mitigating the overfitting issue associated with easily fitted points, in contrast to uniform noise addition.

In summary, the contributions are as follows:
\begin{itemize}
\item We reveal the importance of data augmentation in the context of \textbf{s}tock forecasting and explore the use of \textbf{diff}usion \textbf{s}tock trans\textbf{former} (DiffsFormer for short) to address data scarcity.

\item To adapt DM from generative task to supervised-learning task, we propose to employ ground truth (\eg return ratio) as the conditioning to enhance the relationship between factors and label. Furthermore, we enhance the flexibility of the guidance by integrating a predictor-free guidance approach.

\item Due to the low SNR of factors, we propose to edit the existing samples in a transfer learning manner instead of synthesizing new ones. 
One advantage of this mechanism lies in its capacity to distill new knowledge in contrast to methods like random noise augmentation.

\item We verify the effectiveness of DiffsFormer augmented training in CSI300 and CSI800 with eight commonly used machine learning models. The proposed method achieves 7.2\% and 27.8\% relative improvements in annualized return ratio respectively.
\end{itemize}


%% file: chapters/2_related.tex
\section{Related Works}
\label{sec:related}
In this section, we will introduce the related work in stock forecasting and discuss some time-series diffusion models.

\subsection{Stock Forecasting}
Stock forecasting is a field that utilizes historical time-series data to predict future stock prices. Machine learning models, particularly time-series models such as LSTM, GRU, and Bi-LSTM, have gained popularity in this domain \cite{zou2022stock, DBLP:journals/neco/HochreiterS97, DBLP:journals/corr/ChungGCB14, DBLP:journals/nn/GravesS05}.

Researchers have proposed tailored models to better fit the financial scenario. For example, Li et al. \cite{DBLP:conf/acml/LiSZ18} introduce extra input gates to extract positive and negative correlations between factors. Ding et al. \cite{DBLP:journals/mlc/DingQ20} propose a novel LSTM model to simultaneously predict the opening, lowest, and highest prices of a stock. Agarwal et al. \cite{DBLP:journals/eswa/RatherAS15} propose a hybrid prediction model (HPM) that combines three time-series models. Zhang et al. \cite{DBLP:conf/kdd/ZhangAQ17} propose a State Frequency Memory (SFM) network that decomposes the hidden states of memory cells into multiple frequency components to model different latent trading patterns. Feng et al. \cite{DBLP:conf/ijcai/FengC0DSC19} incorporate a temporal attentive aggregation layer and adversarial training into an LSTM variant. Chen et al. \cite{DBLP:journals/corr/abs-1910-05078} use Bi-LSTM to encode stock data and financial news representations in their Structured Stock Prediction Model (SSPM) and multi-task Structured Stock Prediction Model (MSSPM).

CNNs are also believed to capture important features for predicting stock fluctuations. For instance, Deng et al. \cite{DBLP:conf/www/DengZZCPC19} propose the Knowledge-Driven Temporal Convolutional Network (KDTCN), which integrates knowledge graphs with CNNs to fully utilize industrial relations. Lu et al. \cite{DBLP:journals/nca/LuLWQ21} enhance a CNN-based model by extracting historical influential stock fluctuations with attention mechanism. Chandar \cite{DBLP:journals/ase/Chandar22} transforms technical indicators into images and used them as input for a CNN model.

To handle non-Euclidean structured data, some researchers have incorporated Graph Neural Networks (GNNs) into stock forecasting. Velickovic et al. \cite{DBLP:conf/iclr/VelickovicCCRLB18} construct a graph with stocks as nodes and used graph attention network (GAT) to aggregate neighbor embeddings. Xu et al. \cite{DBLP:journals/isci/XuHYGLZXZL22} construct a stock market relationship graph and extracted information hierarchically. Li et al. \cite{DBLP:conf/ijcai/0101BHCXS20} propose an LSTM Relational Graph Convolutional Network (LSTM-RGCN) model that handles both positive and negative correlations among stocks.

The Transformer model \cite{DBLP:conf/nips/VaswaniSPUJGKP17}, with self-attention and positional encoding mechanisms, has shown great potential in stock forecasting. Ding et al. \cite{DBLP:conf/ijcai/DingWSGG20} improve the Transformer by incorporating multi-scale Gaussian prior, optimizing locality, and implementing Orthogonal Regularization. Yoo et al. \cite{DBLP:conf/kdd/YooSPK21} propose a Data-axis Transformer with Multi-Level Contexts (DTML) to learn the correlations between stocks. Yang et al. \cite{DBLP:conf/www/YangNSD20} introduce the Hierarchical, Transformer-based, multi-task (HTML) model for predicting short-term and long-term asset volatility. For more related works, please refer to \cite{zou2022stock}.

\subsection{Time-series Diffusion Models}
Diffusion models (DMs) are a family of generative models based on deep learning. DMs have demonstrated impressive performance in generating data samples that adhere to the observed data distribution, thus making them a powerful tool for data augmentation. In recent years, DMs have been implemented for multivariate time series (MVTS) applications.

The first prominent model in this domain is TimeGrad, proposed by Rasul et al. \cite{DBLP:conf/icml/RasulSSV21}. TimeGrad builds upon Denoising Diffusion Probabilistic Models (DDPM) and consists of a forward process and a reverse process that are conditioned on an RNN module, which encodes historical information. Yan et al. \cite{DBLP:journals/corr/abs-2106-10121} extend TimeGrad from discrete to continuous by constructing the diffusion process using stochastic differential equations (SDEs). They also discuss the potential of extending RNN to other architectural structures.

More recently, Li et al. \cite{DBLP:conf/nips/LiLWD22} introduce D3VAE to address noisy time series. They designed a novel coupled diffusion process to synthesize features and labels, respectively, and employed a bidirectional variational auto-encoder (BVAE) to bridge the two.

Spatio-temporal graphs (STG) represent a unique form of time-series data that encodes both spatial and temporal interactions among nodes in a graph \cite{DBLP:journals/corr/abs-2301-13629}. Wen et al. \cite{DBLP:journals/corr/abs-2301-13629} introduce DiffSTG, which applies DMs on STG by incorporating a graph convolution layer into the traditional UNet architecture. The model has a DDPM-like structure that is conditioned on both the graph and the temporal information.

For more related works, please refer to \cite{DBLP:journals/corr/abs-2305-00624}.

%% file: chapters/3_background.tex
\section{Background}
\label{sec:background}
In this section, we will introduce some definitions in our work and the problem of stock price forecasting.
\subsection{Problem Formulation}
\label{sec:formulate}
\noindent \textbf{Definition 1. Stock Factors.} Factors are attributes of a stock that are identified as potential drivers of return.

\noindent \textbf{Remark.} To provide clarity, stock factors can be classified into two types: Alpha factors and Beta factors. Alpha factors measure excess returns, while Beta factors measure volatility. Representative examples of these factors include stock volume, value (\ie open, high, close, low prices), and momentum.

\vspace{5pt}
\noindent \textbf{Definition 2. Return Ratio (RR).} The primary objective of stock forecasting is to achieve substantial profits. Previous study \cite{zou2022stock} treat RR as a metric to measure the model performance. RR serves as a crucial indicator to assess the success of stock forecasting models in achieving profitable investment outcomes. Following this setting, we define return ratio as:
\begin{equation}
\label{eq:rr}
   RR(i) = \frac{{Cprice_{t+i}}- Cprice_{t}}{Cprice_{t}},
\end{equation}
where $t$ is the current time, and $i$ denotes the time interval in days. $Cprice_{t}$ denotes the current close price of the stock, and $Cprice_{t+i}$ represents the close price of the same stock after $i$ days. Here, we calculate the return ratio on a daily basis, and often set $i$ to be 5.

\vspace{5pt}
\noindent \textbf{Definition 3. Stock Forecasting.} Based on specific predefined stock factors, our goal is to train a time-series regressor that can accurately predict the future stock behavior of the test stocks in the next few days.

\subsection{Denoising Diffusion Probabilistic Model}
\label{sec:diffusion}
Denoising Diffusion Probabilistic Models (DDPM) have achieved impressive performance in various domains, especially in text-to-image scenarios \cite{nichol2021glide, ramesh2022hierarchical}. Typically, training a DM needs diffusion and denoising processes.

\vspace{5pt}
\noindent \textbf{Diffusion process.} Given a data point $\Mat{x}_{0} \sim q(x_{0})$, the diffusion process gradually adds noise to construct a sequence of step-dependent variables $\{\Mat{x}_{t}\}_{t=1}^{T}$ \cite{wang2023diffusion}. More specifically, it forms a Markov chain as \cite{tashiro2021csdi}:
\begin{equation}
\label{eq:forward}
    q(\Mat{x}_{1:T} | \Mat{x}_{0}) = \prod_{t=1}^{T} q(\Mat{x}_{t} | \Mat{x}_{t-1}), 
\end{equation}
where $q(\Mat{x}_{t} | \Mat{x}_{t-1}) = \mathcal{N}(\Mat{x}_{t}; \sqrt{\alpha_{t}}\Mat{x}_{t-1}, \beta_{t}\Mat{I})$. $\mathcal{N}$ denotes the Gaussian distribution, $\alpha_{t}$ controls the strength of signal retention, and $\beta_{t}$ controls the scale of the added noise. These two scalars are predefined for each step $t$, and one commonly used setting is the variance preserving process \cite{ho2020denoising} where $\alpha_{t} = 1-\beta_{t}$.

\vspace{5pt}
\noindent \textbf{Denoising process.} The goal of the denoising process is to reconstruct the corresponding noise vector by inverting the transformations performed in the diffusion process. This process is defined by another Markov chain \cite{tashiro2021csdi}:
\begin{equation}
    p_{\theta}(\Mat{x}_{0:T})= p(\Mat{x}_{T})\prod_{t=1}^{T}p_{\theta}(\Mat{x}_{t-1}|\Mat{x}_{t}),
\end{equation}
\noindent where $\Mat{x}_{T} \sim \mathcal{N}(0, \Mat{I})$. $p_{\theta}$ is the distribution estimation of $q$, for which $p_{\theta} (\Mat{x}_{t-1} | \Mat{x}_{t}) = \mathcal{N}(\Mat{x}_{t-1}; \mu_{\theta} (\Mat{x}_{t}, t), \sigma_{\theta} (\Mat{x}_{t}, t)\Mat{I})$. Concretely, for each sample in the batch, a time step $t$ is uniformly sampled from $\{1, 2, ..., T\}$, followed by the adjustment of the noise at time $t$.


\begin{algorithm}[t]
\caption{\textbf{DiffsFormer Training}}\label{alg:training}
\begin{algorithmic}
\STATE{\textbf{Input:} stock data $\Mat{X} \in \mathbb{R}^{n\times d \times k}$, diffusion step $T$}
\FOR{$t=1$ to $T$}
\STATE{initialize $\beta_{t}$ and calculate $\overline{\alpha}_{t}$}
\ENDFOR
\STATE{Select an editing step $T' \leq T$}
\WHILE{Not Converge}
\STATE{$i \sim \rm{Uniform}\{1, 2, \cdots, n\}$}
\STATE{$t \sim \rm{Uniform}\{1, 2, \cdots, T'\}$}
\STATE{$\epsilon \sim \mathcal{N}(0, \Mat{I})$}
\STATE{$\Mat{x}_{0}:= \Mat{X}[i]$}
\STATE{calculate $\Mat{x}_{t}$ given $\Mat{x}_{0}$ with Eq.\eqref{eq:calxt}}
\STATE{calculate $\mathcal{L}_{da}$ with Eq.\eqref{eq:daloss}}
\STATE{Take a gradient step on $\nabla_{\theta}\mathcal{L}_{da}$}
\ENDWHILE
\end{algorithmic}
\end{algorithm}

\vspace{5pt}
\noindent \textbf{Inference process.} Once $\theta$ is well-trained, the DM can generate samples from the standard Gaussian distribution with $\Mat{x}_{T} \sim \mathcal{N}(0, \Mat{I})$ and then repeatedly recover $\Mat{x}_{T} \rightarrow \cdots \rightarrow \Mat{x}_{t} \rightarrow \Mat{x}_{t-1} \rightarrow \cdots \rightarrow \Mat{x}_{0}$ given $p_{\theta}(\Mat{x}_{t-1}|\Mat{x}_{t})$. As $T\rightarrow\infty$, the generative process modeled with Gaussian conditional distributions becomes a good approximation. 

%% file: chapters/4_method.tex
\section{Methodology}
\label{sec:method}

The stock forecasting task is challenging primarily because of the scarcity of data. To harness the full potential of machine learning models, a sufficient amount of high-quality data is crucial. However, obtaining such high-quality stock data for a specific target domain is rare and can often be restricted as commercial secrets. In this work, we utilize the power of DM and introduce a novel approach, DiffsFormer. It generates additional data points and facilitates factor augmentation, enabling us to forecast the likely RR of real-world stocks despite data scarcity.

\begin{algorithm}[t]
\caption{\textbf{DiffsFormer Inference}}\label{alg:sampling}
\begin{algorithmic}
\STATE{\textbf{Input:} number of data to be generated $m$, sampling steps $l$, conditionings $c$ (if guidance is enabled), editing step $T'$ selected during training}
\WHILE{Generated point $<m$}
\STATE{$i \sim \rm{Uniform}\{1, 2, \cdots, n\}$}
\STATE{$\Mat{x}_{0}:= \Mat{X}[i]$}
\STATE{calculate $\Mat{x}_{T'}$ given $\Mat{x}_{0}$ with Eq.\eqref{eq:calxt}}
\STATE{$\Mat{x}_{\tau_{l}} := \Mat{x}_{T'}$}
\FOR{$t=l$ to $0$}
\STATE{calculate $\Mat{x}_{\tau_{t-1}}$ with DDIM sampling}
\ENDFOR
\ENDWHILE
\end{algorithmic}
\end{algorithm}

\subsection{Diffusion-based Data Augmentation}
\label{sec:dda}
Following \cite{ho2020denoising, nichol2021glide}, DiffsFormer contains diffusion and denoising processes like most of the DMs do. 

\vspace{5pt}
\noindent \textbf{Diffusion process.} 
In stock forecasting, the input data $\Mat{X} \in \mathbb{R}^{n \times d \times k}$ consists of $n$ real stocks along with their recent $k$-day historical factors, for which $d$ is the factor dimension. We treat each stock $\Mat{x}$ (\ie a row of $\Mat{X}$) as $\Mat{x}_{0}$ sampled from $q(\Mat{x}_{0})$, and add random noise to perform a transition according to Eq.\eqref{eq:forward}. Thanks to the reparameterization trick \cite{ho2020denoising}, we can obtain the conditional distribution $q(\Mat{x}_{t}|\Mat{x}_{0})$ for each stock \cite{wang2023diffusion, tashiro2021csdi}:

\begin{equation}
    q(\Mat{x}_{t} | \Mat{x}_{0}) = \mathcal{N}(\Mat{x}_{t};\sqrt{\overline{\alpha}_{t}}
\Mat{x}_{0}, (1-\overline{\alpha}_{t})\Mat{I}), 
\label{eq:calxt}
\end{equation}
\noindent where $\overline{\alpha}_{t}=\prod_{i=1}^{t}\alpha_{i}$ and $\alpha_{t}=1-\beta_{t}$. Then, $\Mat{x}_{t}$ is approximated as $\Mat{x}_{t} = \sqrt{\overline{\alpha}_{t}}\Mat{x}_{0}+(\sqrt{1-\overline{\alpha}_{t}})\epsilon$ where $\epsilon \sim \mathcal{N}(0,\Mat{I})$.

\vspace{5pt}
\noindent \textbf{Denoising process.} 
During the denoising process, we subtract noise from $\Mat{x}_{t}$ to recover the corresponding $\hat{\Mat{x}}_{0} \sim q(\Mat{x}_{0})$. As mentioned in \S \ref{sec:diffusion}, we parameterize $p_{\theta} (\Mat{x}_{t-1} | \Mat{x}_{t})$ through a neural network to estimate $q(\Mat{x}_{t-1} | \Mat{x}_{t}, \Mat{x}_{0})$. Specifically, we have $p_{\theta} (\Mat{x}_{t-1} | \Mat{x}_{t})=\mathcal{N}(\Mat{x}_{t-1}; \mu_{\theta} (\Mat{x}_{t}, t), \Mat\Sigma_q(t)\Mat{I})$ with:

\begin{equation}
\label{eq:ddpm_mean_var}
    \begin{aligned}
        \mu_{\theta} (\Mat{x}_{t}, t)&=\frac{1}{\sqrt{\alpha}_{t}}(\Mat{x}-\frac{\beta_{t}}{\sqrt{1-\overline{\alpha}_{t}}}\epsilon_{\theta}(\Mat{x}_{t}, t)) 
        \\
        \Mat\Sigma_q(t)&=\frac{(1-\overline{\alpha}_{t-1})\beta_{t}}{1-\overline{\alpha}_{t}}
        ,
    \end{aligned}
\end{equation}
where $\epsilon_{\theta}(\Mat{x}_{t}, t)$ is the trainable noise term to predict $\epsilon$ in the diffusion process. 

\vspace{5pt}


\vspace{5pt}
\noindent \textbf{Objective.} The overall learning objective is to minimize the error in estimating $\epsilon$ using $\epsilon_{\theta}(\Mat{x}_{t}, t)$ \cite{nichol2021glide}. Formally, we aim to solve the following optimization problem:
\begin{equation}
    \label{eq:daloss}
    \begin{aligned}
    \mathcal{L}_{da}= \min\limits_{\theta} \mathbb{E}_{\Mat{x}_{0} \sim q(\Mat{x}_{0}),\epsilon \sim \mathcal{N}(0,\Mat{I}),t \sim {\rm Uniform}} ||\epsilon - \epsilon_{\theta} &(\Mat{x}_{t}, t)||_{2}^{2} \\
    s.t. ~ \Mat{x}_{t} =\sqrt{\overline{\alpha}_{t}}\Mat{x}_{0}+(\sqrt{1-\overline{\alpha}_{t}}&)\epsilon.
    \end{aligned}
\end{equation}

\begin{figure}[t]
\centerline{\includegraphics[width=0.5\textwidth]{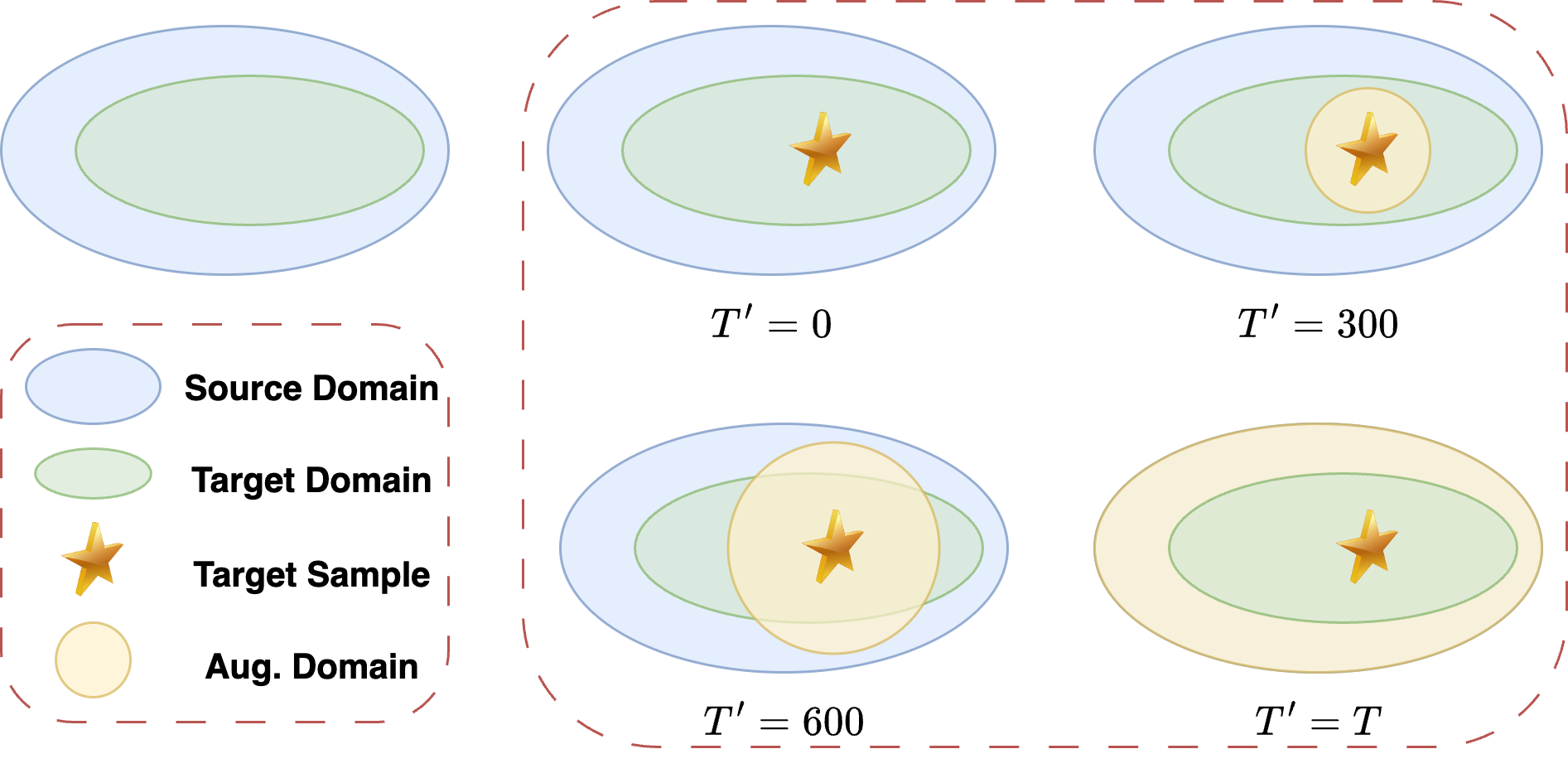}}
\caption{Illustration of the editing step.}
\vspace{-10pt}
\label{fig:tfunction}
\end{figure}

\noindent \textbf{Inference acceleration.} 
In DDPM, the lack of parallelism during the transition of DMs leads to slow inference. To tackle this problem,  DDIM \cite{song2020denoising} accelerates samplings by modifying the 
forward process as a non-Markov process, i.e.,
\begin{equation}
q_\sigma\left({\Mat x}_{1: T} \mid {\Mat x}_0\right) =q_\sigma\left({\Mat x}_T \mid {\Mat x}_0\right) \prod_{t=2}^T q_\sigma\left({\Mat x}_{t-1} \mid {\Mat x}_t, {\Mat x}_0\right),
\end{equation}
where $q_\sigma\left({\Mat x}_{t-1} \mid {\Mat x}_t, {\Mat x}_0\right)$ is parameterized by $\sigma$ which means the magnitude of the stochastic process. When setting $\sigma_t = \Mat\Sigma_q(t)$ for all steps, the forward process collapses to Markovian and the denosing process becomes the same as shown in Eq.(\ref{eq:ddpm_mean_var}). Specifically, when setting $\sigma_t = 0$, the corresponding denosing process becomes deterministic and thus sampling could be accelerated along the deterministic path.
Technically, we follow the deterministic sampling design and create $\{\tau_{i}\},\{i=1 \cdots l\}$ as a sub-sequence of $\{t=1, 2, \cdots, T\}$, where $l$ is the length of the sub-sequence. The denoising process can now be completed in just $l \ll T$ steps armed with DDIM sampling.

\begin{figure}[t]
\centerline{\includegraphics[width=0.5\textwidth]{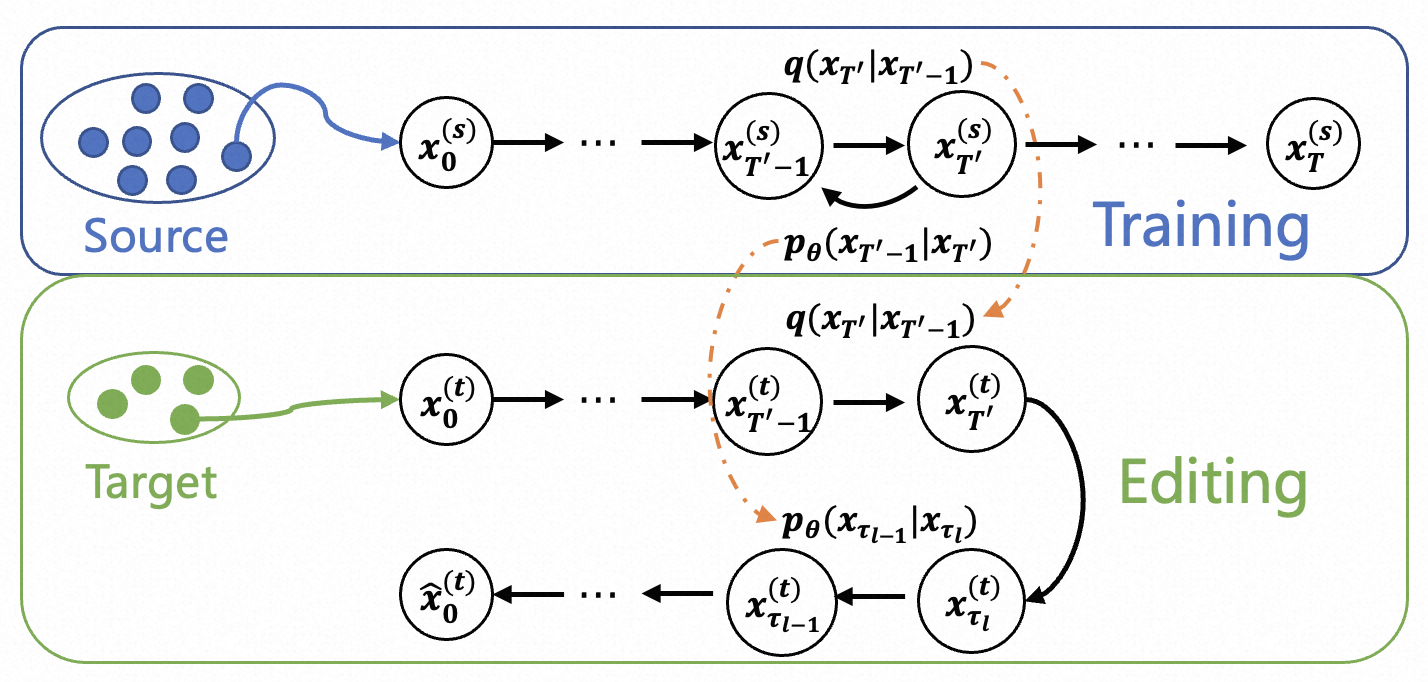}}
\caption{The training and the editing topology.}
\vspace{-15pt}
\label{fig:ddpm_topo}
\end{figure}

\noindent \textbf{Factor editing with transfer learning.} To alleviate data homogeneity issue, we augment the raw factors in target domain by going through a noising-denoisng process. Instead of generating synthetic factors from pure noise which hardly ensures data fidelity, we adopt a different approach by editing the original factors rather than generating entirely new ones. Moreover, due to the intrinsic low SNR nature of the factors, we design a transfer learning framework to distill new knowledge and information into edited data from a larger, different domain. Concretely, DiffsFormer of diffusion step $T$ is first trained on the source domain $\Mat{X}^{(s)}$. During the inference process, we begin with a data point in the target domain $\Mat{x}_{0}^{(t)}$, corrupt it for $T' \ll T$ steps to get a seed point: $\Mat{x}_{0}^{(t)} \rightarrow \Mat{x}_{1}^{(t)} \rightarrow \cdots \rightarrow \Mat{x}_{T'}^{(t)}$. Then, we reverse the process from the seed to obtain a new data point $\hat{\Mat{x}}_{T'}^{(t)}$ in the target domain: $\Mat{x}_{T'}^{(t)} \rightarrow \Mat{x}_{T'-1}^{(t)} \rightarrow \cdots \rightarrow \hat{\Mat{x}}_{0}^{(t)}$. Since the target domain is a subset of the source domain, this procedure distills new knowledge and information and enhances the data heterogeneity. Moreover, since the inference process starts from the seed, we can successfully edit existing samples. As illustrated in Figure \ref{fig:tfunction}, $T'$ can control the strength of knowledge distillation: a larger $T'$ makes the generated data resemble the feature distribution of the source domain more closely, while a smaller $T'$ makes the generated data closer to the target domain data $\Mat{x}_{0}^{(t)}$. We term $T'$ as the editing step. By doing so, we improve the fidelity of the generated data, avoiding creating data from pure noise. Recent works \cite{DBLP:journals/corr/abs-2210-07574} term this trick as ``real guidance''. An illustration of the process is shown in Figure \ref{fig:ddpm_topo}.

\vspace{5pt}
\noindent \textbf{Time efficiency improvements.} In our transfer learning framework, it is obvious that there is no need to optimize $\epsilon_{\theta}(\Mat{x}_{t}, t)$ for $t > T'$. On top of that, we develop a trick to speed up the training of the framework. Concretely, we initialize $\alpha$ and $\beta$ with the total diffusion steps $T$; but during training, we sample training steps $t$ from ${\rm Uniform} \{1, 2, \cdots, T'\}$ instead of ${\rm Uniform} \{1, 2, \cdots, T\}$. We discover that this modification results in a sharper loss curve during training as shown in Figure \ref{fig:smoothed-loss}. The detailed algorithms for training and inference are shown in Algorithms \ref{alg:training} and \ref{alg:sampling}, respectively.

\subsection{Conditional Diffusion Augmentation}
\begin{figure}[t]
\centerline{\includegraphics[width=0.5\textwidth]{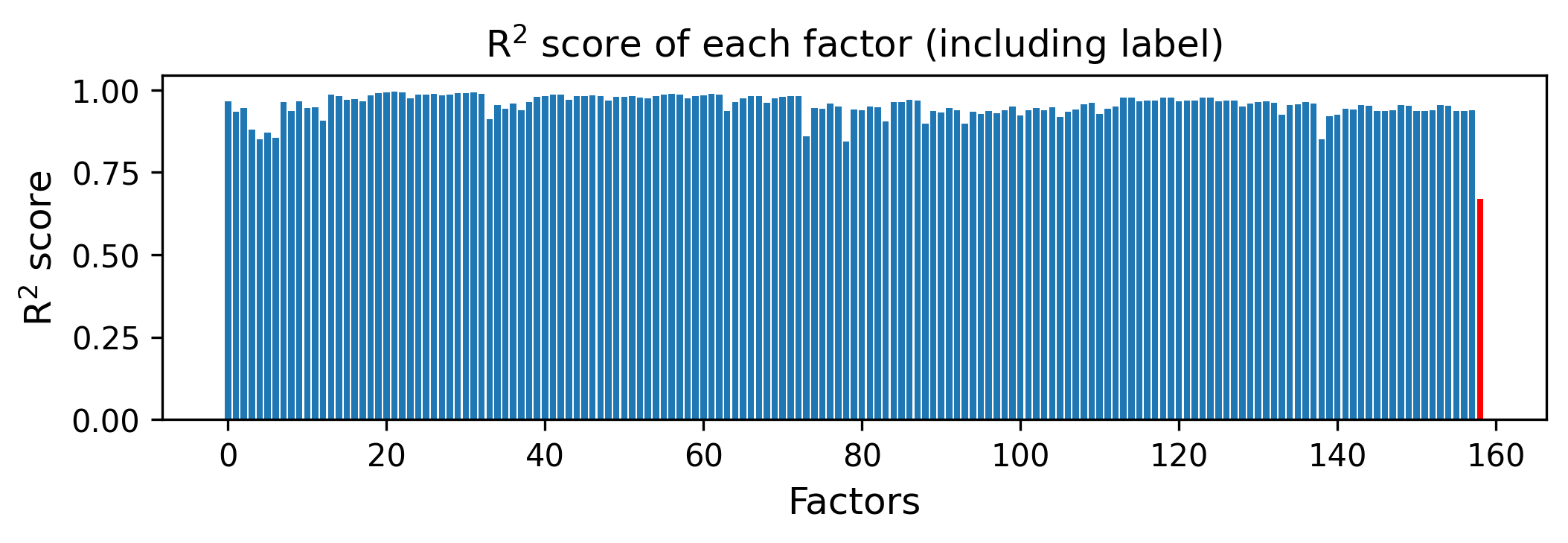}}
\caption{The $R^{2}$ score between the generated and the original factors and label. $R^{2}$ score is the square of the Pearson Correlation. The blue bars represent the $R^{2}$ scores of 158 factors, while the red bar shows the $R^{2}$ score of the label.}
\vspace{-10pt}
\label{r_score}
\end{figure}

Most generative tasks do not have the demands for label generation. However, in the stock forecasting task, a clean and informative supervised signal is essential for training the regressor. Directly appending the label to the factor vector is ineffective. Taking Figure \ref{r_score} as an example, we calculate the $R^{2}$ score between the augmented and the original factors (label). It is observed that the label has the least correlation among the 159 dimensions. Furthermore, we report an experimental comparison between the settings of label-generation and label-conditioning in Table \ref{tab:label-gene}. We suppose that generated label fails to serve as the accurate supervised signal for the generated feature. As an alternative, we pave the way to control the synthesis process through guidance inputs, including labels and industry information \cite{rombach2022high}. At the meantime, we keep the labels of the generated features the same as those of the original features to adapt DMs from the generative task to the regression task.

\begin{table}[t]
\setlength{\tabcolsep}{4mm}{
\caption{Model performance with label-generation and label-condition mechanisms}
\label{tab:label-gene}
\begin{tabular}{c|c|c}
\toprule
            & Label-generation   &  Label-condition     \\ \midrule
Model Performance & 0.159327 & 0.312679 \\ \bottomrule
\end{tabular}%
}
\vspace{-10pt}
\end{table}

\subsubsection{Predictor Guidance}
One straightforward approach is to introduce an auxiliary classifier $\phi$ that estimates $p_{\phi}(y|\Mat{x}_{t}, t)$ \cite{song2020score, dhariwal2021diffusion}. The gradients $\nabla_{\Mat{x}_{t}}{\rm log}p_{\phi}(y|\Mat{x}_{t}, t)$ obtained during training guide the synthesis of data points in the inference process. 
Concretely, $\epsilon_{\theta}(\Mat{x}_{t}, t)$ in Eq.\eqref{eq:daloss} becomes:
\begin{equation}
\label{eq:classguidance}
    \overline{\epsilon}_{\theta} (\Mat{x}_{t}, t, y) = {\epsilon}_{\theta} (\Mat{x}_{t}, t) - \sqrt{1-\overline{\alpha}_{t}}\omega\nabla_{\Mat{x}_{t}}{\rm log}p_{\phi}(y|\Mat{x}_{t}, t),
\end{equation}
\noindent where $\omega$ is a scalar to control the guidance strength. 

This guiding format is widely used to introduce new concepts in image synthesis \cite{song2020score, dhariwal2021diffusion}. However, in our work, the target label is continuous, which is largely unexplored to the best of our knowledge. One of the differences between continuous and discrete label space is the use of time step $t$. Note that in discrete space, a noisy image may belong to an unknown class, especially when it is pure noise. Therefore, the classifier takes time step $t$ as input to incorporate diffusion information. In contrast, in our work, every sequence is meaningful since the expected label is continuous. On top of that, we refine Eq.\eqref{eq:classguidance} by decoupling the training of the predictor from DMs and ignoring time step $t$ in the input:
\begin{equation}
\label{eq:regressionguidance}
    \overline{\epsilon}_{\theta} (\Mat{x}_{t}, t, y) = {\epsilon}_{\theta} (\Mat{x}_{t}, t) - \sqrt{1-\overline{\alpha}_{t}}\omega\nabla_{\Mat{x}_{t}}(p_{\phi^{*}}(\Mat{x}_{t})-y)^{2},
\end{equation}
where $\phi^{*}$ is the pre-trained separate predictor whose parameters remain fixed during the inference. Similar to other modules, we choose the Transformer as the auxiliary predictor for its excellent performance in sequence tasks. One of the advantages of the predictor guidance mechanism is that it maintains the plug-and-play design of the framework, making our model scalable.

\begin{figure*}[t]
\centerline{\includegraphics[width=\textwidth]{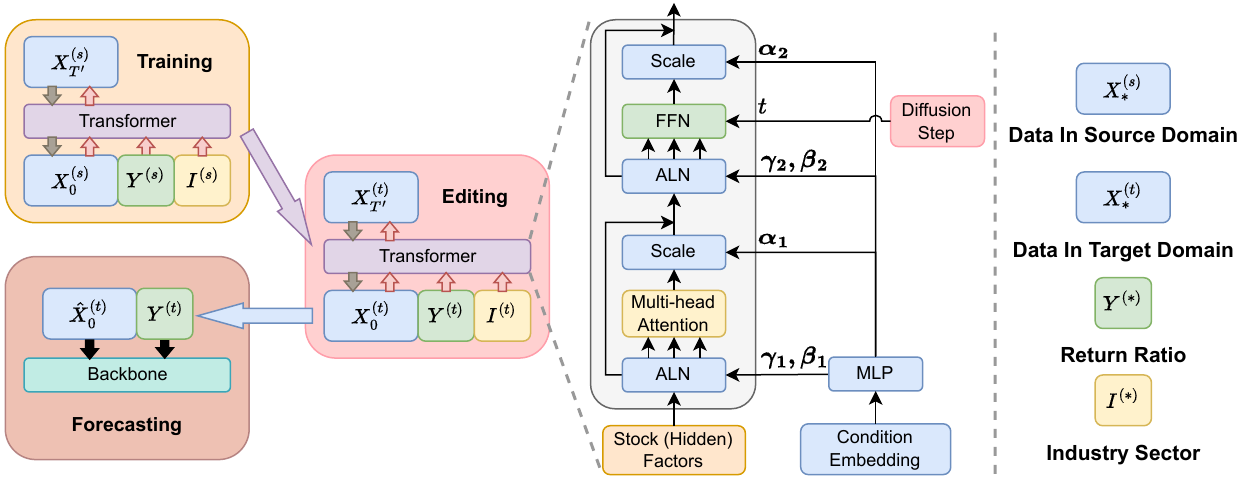}}
\caption{DiffsFormer overview. Y denotes label and I denotes industry sector. DiffsFormer is initially trained on a large-scale source domain, incorporating conditional guidance to ensure improved model performance. When presented with a specific downstream task, we employ DiffsFormer to augment the training procedure by editing existing samples.}
\vspace{-10pt}
\label{fig:topo}
\end{figure*}

\subsubsection{Predictor-free Guidance}
Besides predictor guidance, an alternate approach is the predictor-free guidance. Due to the fact that the additional predictor complicates the training pipeline, some research work explore to eliminate it \cite{ho2022classifier, peebles2022scalable, nichol2021glide}. Moreover, fusing the obtained gradient into DM training can be considered a gradient-based adversarial attack \cite{ho2022classifier}, making the performance of predictor guidance poorer than that of the predictor-free guidance in most cases \cite{peebles2022scalable, nichol2021glide}. Therefore, in our work, we also explore the implementation of the predictor-free guidance mechanism.

Upon closer inspection of Eq.\eqref{eq:regressionguidance}, we observe that predictor guidance works by changing the sampling distribution from $p_{\theta}(\Mat{x}_{t}|y)$ to $\overline{p}_{\theta}(\Mat{x}_{t}|y)$ where:
\begin{equation}
    \overline{p}_{\theta}(\Mat{x}_{t}|y) \propto p_{\theta}(\Mat{x}_{t}|y)p_{\theta}(y|\Mat{x}_{t})^{\omega}.
\end{equation}
As the guidance strength $\omega$ gets larger, DM receives more rewards when generating $\Mat{x}_{t}$ for which the auxiliary predictor assigns a high probability $p_{\theta}(y|\Mat{x}_{t})$ to the ground truth $y$. According to \cite{ho2022classifier}, the same effect can be achieved by jointly training conditional and unconditional DMs. Specifically, the inference process is performed in the form of:
\begin{equation}
    \hat{\epsilon}_{\theta} (\Mat{x}_{t}, c) = {\epsilon}_{\theta} (\Mat{x}_{t}, \emptyset) + \omega_{free} \cdot (\epsilon_{\theta}(\Mat{x}_{t}, c)-\epsilon_{\theta}(\Mat{x}_{t}, \emptyset)),
\end{equation}
\noindent where $c$ denotes the condition vectors and $\emptyset$ denotes a learnable null vector. During training, $c$ is randomly replaced with $\emptyset$ with a fixed probability to train an unconditional DM. Note that $\omega_{free}$ shall be greater than $1$ in predictor-free guidance. The advantages of predictor-free guidance are: 1) it is a simple approach since no auxiliary predictor is needed; 2) it is flexible since it supports other types of conditionings beyond return-ratio labels. In our work, we further explore the use of industry information. We observe that stocks in different industries tend to perform in different patterns. For instance, financial stocks (\eg banks) usually have low yields but enjoy low volatility, while many information technology stocks have high yields but undertake high volatility. Furthermore, we can synthesize industry-specific data to improve model performance in specific industry sector. One of the unappealing properties of the predictor-free guidance is that it injects the conditionings during the training of DMs. As a result, when adding or modifying conditionings, we need to retrain the DMs although it is time-consuming.

\subsection{Model Details}

\subsubsection{Model Overview}
Figure \ref{fig:topo} elucidates the overall framework of our stock price forecasting model. Generally, it consists of a conditioned transformer-based data augmentation module and a backbone return ratio predictor. The framework is designed with several considerations: 1) DMs acts as a plug-and-play data augmentation module, so it can be deployed to different backbones without retraining; 2) our data is organized in sequences, so we explore the use of transformers to better capture the autocorrelation in the sequence, as opposed to the commonly adopted UNet \cite{DBLP:conf/miccai/RonnebergerFB15} in text-to-image generation models; 3) the transfer-based editing framework distills new knowledge and information, meanwhile prevents the new data copy from deviating from the original data too much.

\subsubsection{Architecture Improvement} 
Following \cite{peebles2022scalable}, we adjust the transformer structure to fit predictor-free guidance. Specifically, as shown in Figure \ref{fig:topo}, we feed the diffusion steps into the transformer module. Furthermore, it contains an adaptive layer norm module and a zero-initialized scalar module.

\vspace{5pt}
\noindent \textbf{Time step encoding.} During the training, DiffsFormer needs to know which diffusion step it is trained for. In our work, we encode the current diffusion step into \textit{Sinusoidal positional encoding} and add it to the feed forward network. This embedding scheme is appropriate to encode the time step information, the positional encoding at position $p+k$ can be linearly represented by the positional encoding at position $p$. 

\vspace{5pt}
\noindent \textbf{Adaptive layer norm (ALN).} Adaptive layer norm \cite{DBLP:conf/aaai/PerezSVDC18} are widely adopted in generative models \cite{dhariwal2021diffusion, DBLP:conf/iclr/BrockDS19}. In the transformer block, we append an additional ALN layer after the standard layer norm. Notably, ALN regresses the scale and shift parameters $\bm \gamma$ and $\bm \beta$ from two affine transformations $f$ and $h$ to the sum of conditioning vectors.

\vspace{5pt}
\noindent \textbf{Zero initialization.} In addition to ALN layer, many DMs \cite{dhariwal2021diffusion, ho2020denoising, peebles2022scalable} also incorporate zero initialization in the framework, meaning that the model parameters are initialized as zero such that the conditioning is ineffective when training just starts. In this case, MLP is initialized to make $\bm \alpha_1$ and $\bm \alpha_2$ equal to $0$, and thus transformer block becomes an identity function \cite{peebles2022scalable,DBLP:journals/corr/GoyalDGNWKTJH17} (\ie input tokens are directly fed to the next layer).

%% file: chapters/5_experiment.tex
\section{Experiment}
\label{sec:experiment}
In this section, we conduct experiments on the real-world stock data from 2008 to 2022 provided by \cite{DBLP:journals/corr/abs-2009-11189} with the aim of answering the following research questions:

\begin{itemize}
    \item \textbf{RQ1:} Is proposed DiffsFormer flexible to various backbones and helpful in enhancing them?
    \item \textbf{RQ2:} Can DiffsFormer alleviate data scarcity? How effective are the proposed components?
    \item \textbf{RQ3:} How does DiffsFormer help control the volatility of the backbone models?
\end{itemize}

\subsection{Dataset}
\subsubsection{Dataset} Following \cite{DBLP:journals/corr/abs-2110-13716, DBLP:journals/corr/abs-2212-08656}, we evaluate the proposed framework on two real-world stock datasets: CSI 300 and CSI 800. The CSI 300 comprise the largest 300 stocks traded on the Shanghai Stock Exchange and the Shenzhen Stock Exchange\footnote{https://en.wikipedia.org/wiki/CSI\_300\_Index}, and represents the performance of the whole A-share market in China. 
CSI 800 is a larger dataset consisting of CSI 500 and CSI 300, aiming to add some stocks with smaller size. Note that DiffsFormer aims at editing the existing samples with new information from a larger domain. Hence in practice, we use all stocks in the China A-share market to train the DM and editing on CSI 300 and CSI 800, respectively.

\subsubsection{Factors} We use the Alpha158 factors provided by the AI-oriented quantitative investment platform Qlib\footnote{https://github.com/microsoft/qlib}. These factors review the basic stock information including \textit{kbar, price, volume, and some rolling factors} in different time windows. For each stock at date $t$, we look back 8 days to construct a sequence of factor as $\Mat{x} \in \mathbb{R}^{8\times158}$. During the time span between 2008-01-01 and 2022-09-30, the number of sequences is 2109804. Hence our input matrix $\Mat{X}$ is of shape 2109804$\times$8$\times$158.

\begin{table}[t]
\setlength{\tabcolsep}{4.5mm}{
\caption{Hyper-parameters and the search range, the optimal parameters are indicated in boldface.}
\label{tab:para}
\begin{tabular}{c|c}
\toprule
Parameters                    & Search Range                   \\\midrule
editing step during inference & \{200, 300, \textbf{400}, 500\}         \\
layers in DM     & \{3, \textbf{6}\}                       \\
stop loss thred               & \{0.6, 0.8, 0.9, 0.95, \textbf{0.965}, 1\} \\
batch norm                    & \{\textbf{False}, True\}                \\ 
norm first                    & \{False, \textbf{True}\}                \\ 
guidance strength             & \{1.1, 2, \textbf{3}, 4\}               \\ 
industry condition            & \{False, \textbf{True}\}                \\
label condition               & \{False, \textbf{True}\}                \\ 
market value condition        & \{\textbf{False}, True\}                \\ 
market value groups           & \{3, \textbf{10}\}                      \\ \bottomrule
\end{tabular}%
}
\vspace{-10pt}
\end{table}

\subsection{Reproducibility}
In this subsection, we introduce some details of the proposed work for easier reproduction.

\begin{table*}[t]
\setlength{\tabcolsep}{1.5mm}{
\caption{Performance comparison on CSI300. The better results are indicated in boldface.}
\label{tab:csi300}
\begin{tabular}{c|cccccccccc|cccccccccc}
\toprule
& \multicolumn{9}{c}{\textbf{CSI300}}                          \\ 
\cmidrule(l){2-10} 
\textbf{Methods} & \multicolumn{3}{c|}{\textbf{RR}}         & \multicolumn{3}{c|}{\textbf{IC}} & \multicolumn{3}{c}{\textbf{RankIC}} \\ 
                        & original & augmentation & \multicolumn{1}{c|}{improvements} & original    & augmentation     & \multicolumn{1}{c|}{improvements} & original    & augmentation      & \multicolumn{1}{c}{improvements}   \\ 
                        \midrule
\textbf{MLP}                         & 0.2093\tiny{±0.0300} & \textbf{0.2163\tiny{±0.0210}} & \multicolumn{1}{c|}{3.34\%} & 0.0508\tiny{±0.0044}      &    \textbf{0.0537\tiny{±0.0026}} & \multicolumn{1}{c|}{5.71\%}  & 0.0499\tiny{±0.0059} & \textbf{0.0509\tiny{±0.0034}} & \multicolumn{1}{c}{2.00\%}   \\ 
\textbf{LSTM}                         & 0.2312\tiny{±0.0308} & \textbf{0.2336\tiny{±0.0219}} & \multicolumn{1}{c|}{1.04\%} & \textbf{0.0516\tiny{±0.0022}}      &   0.0429\tiny{±0.0026}    &\multicolumn{1}{c|}{-16.86\%} & \textbf{0.0519\tiny{±0.0021}} & 0.0455\tiny{±0.0021}   & \multicolumn{1}{c}{-12.33\%}   \\ 
\textbf{GRU}                        & 0.2161\tiny{±0.0293} & \textbf{0.2413\tiny{±0.0149}} & \multicolumn{1}{c|}{11.66\%} & \textbf{0.0536\tiny{±0.0038}}      & 0.0511\tiny{±0.0012}     &\multicolumn{1}{c|}{-4.66\%} & \textbf{0.0552\tiny{±0.0037}} & 0.0516\tiny{±0.0012} & \multicolumn{1}{c}{-6.52\%}  \\
\textbf{SFM}                        & 0.2189\tiny{±0.0325} & \textbf{0.2200\tiny{±0.0175}} & \multicolumn{1}{c|}{0.50\%} & 0.0505\tiny{±0.0018}      & \textbf{0.0510\tiny{±0.0025}}     &\multicolumn{1}{c|}{0.99\%} & 0.0507\tiny{±0.0026} & \textbf{0.0526\tiny{±0.0029}} & \multicolumn{1}{c}{3.75\%}  \\
\textbf{GAT}                        & 0.2461\tiny{±0.0176} & \textbf{0.2701\tiny{±0.0168}} & \multicolumn{1}{c|}{9.75\%} & \textbf{0.0558\tiny{±0.0012}}      &  0.0532\tiny{±0.0007}      &\multicolumn{1}{c|}{-4.66\%} & 0.0540\tiny{±0.0014} & \textbf{0.0551\tiny{±0.0006}} & \multicolumn{1}{c}{2.04\%}  \\
\textbf{ALSTM}                      & 0.2047\tiny{±0.0351} & \textbf{0.2317\tiny{±0.0233}} & \multicolumn{1}{c|}{13.19\%} & \textbf{0.0502\tiny{±0.0027}}  & 0.0450\tiny{±0.0023} & \multicolumn{1}{c|}{-10.36\%}  & \textbf{0.0510\tiny{±0.0031}} & 0.0439\tiny{±0.0019} & -13.92\%  \\

\textbf{HIST}                       & 0.2272\tiny{±0.0352} & \textbf{0.2410\tiny{±0.0207}} & \multicolumn{1}{c|}{6.07\%} & \textbf{0.0547\tiny{±0.0011}}  & 0.0518\tiny{±0.0032}      & \multicolumn{1}{c|}{-5.30\%}  & \textbf{0.0545\tiny{±0.0023}} & 0.0535\tiny{±0.0025} &  -1.83\% \\

\textbf{Transformer}                & 0.2789\tiny{±0.0376} & \textbf{0.3127\tiny{±0.0113}} & \multicolumn{1}{c|}{12.12\%} & 0.0598\tiny{±0.0031}  & \textbf{0.0603\tiny{±0.0025}}    & \multicolumn{1}{c|}{0.83\%}  & 0.0638\tiny{±0.0024} & \textbf{0.0672\tiny{±0.0017}} &   5.33\% \\

\bottomrule
\end{tabular}%
}
\end{table*}

\begin{table*}[t]
\setlength{\tabcolsep}{1.5mm}{
\caption{Performance comparison on CSI800. The better results are indicated in boldface.}
\label{tab:csi800}
\begin{tabular}{c|cccccccccc|cccccccccc}
\toprule
 & \multicolumn{9}{c}{\textbf{CSI800}}                                                                              \\ \cmidrule(l){2-10} 
\textbf{Methods}                         & \multicolumn{3}{c|}{\textbf{RR}}         & \multicolumn{3}{c|}{\textbf{IC}} & \multicolumn{3}{c}{\textbf{RankIC}} \\ 
                         & original & augmentation & \multicolumn{1}{c|}{improvements} & original    & augmentation     & \multicolumn{1}{c|}{improvements} & original    & augmentation      & \multicolumn{1}{c}{improvements}   \\ \midrule
\textbf{MLP}                         & 0.1037\tiny{±0.0383} & \textbf{0.1161\tiny{±0.0223}}& \multicolumn{1}{c|}{11.96\%} & 0.0386\tiny{±0.0023}      &  \textbf{0.0399\tiny{±0.0006}}    &\multicolumn{1}{c|}{3.37\%} & 0.0450\tiny{±0.0048} & \textbf{0.0467\tiny{±0.0035}}& \multicolumn{1}{c}{3.78\%}   \\ 
\textbf{LSTM}                         & 0.1248\tiny{±0.0282} & \textbf{0.1298\tiny{±0.0317}} & \multicolumn{1}{c|}{4.01\%} & 0.0377\tiny{±0.0017}      &  \textbf{0.0412\tiny{±0.0008}}      &\multicolumn{1}{c|}{9.28\%} & \textbf{0.0500\tiny{±0.0030}} & 0.0494\tiny{±0.0010} & \multicolumn{1}{c}{-1.20\%}   \\ 
\textbf{GRU}                        & 0.0758\tiny{±0.0307} & \textbf{0.1295\tiny{±0.0292}} & \multicolumn{1}{c|}{70.84\%} & \textbf{0.0380\tiny{±0.0026}}      & 0.0376\tiny{±0.0010}     & \multicolumn{1}{c|}{-1.05\%} & 0.0493\tiny{±0.0030} & \textbf{0.0511\tiny{±0.0011}} & 3.65\%\\

\textbf{SFM}                        & 0.0906\tiny{±0.0413} & \textbf{0.1250\tiny{±0.0375}} & \multicolumn{1}{c|}{37.97\%} & \textbf{0.0385\tiny{±0.0005}}      & 0.0365\tiny{±0.0015} & \multicolumn{1}{c|}{-5.19\%}  & 0.0485\tiny{±0.0011} & \textbf{0.0487\tiny{±0.0022}} &  0.41\% \\
\textbf{GAT}                        & 0.1814\tiny{±0.0309} & \textbf{0.2013\tiny{±0.0210}} & \multicolumn{1}{c|}{10.97\%} & 0.0379\tiny{±0.0005}      & \textbf{0.0397\tiny{±0.0003}}      & \multicolumn{1}{c|}{4,75\%}  & 0.0483\tiny{±0.0009} & 0.0483\tiny{±0.0009} & 0.00\% \\
\textbf{ALSTM}                        & 0.1030\tiny{±0.0253} & \textbf{0.1518\tiny{±0.0290}} & \multicolumn{1}{c|}{50.29\%} & 0.0316\tiny{±0.0031}      & \textbf{0.0383\tiny{±0.0013}}   & \multicolumn{1}{c|}{21.20\%}  & 0.0418\tiny{±0.0034} & \textbf{0.0492\tiny{±0.0015}} & 17.70\%\\
\textbf{Transformer}                        & 0.1751\tiny{±0.0386} & \textbf{0.1903\tiny{±0.0382}} & \multicolumn{1}{c|}{8.68\%} &  0.0423\tiny{±0.0028}   & \textbf{0.0426\tiny{±0.0018}}      & \multicolumn{1}{c|}{0.71\%} & \textbf{0.0573\tiny{±0.0016}} & 0.0556\tiny{±0.0022} & -2.97\%\\

\bottomrule
\end{tabular}%
}
\vspace{-10pt}
\end{table*}
\subsubsection{Robust Z-score Normalization} Generally, the values between factors are not in the same scale. To address this issue, we adopt \textit{Robust Z-score Normalization} within stocks. Based on z-score, robust z-score replace mean and standard deviation with median (MED) and the median absolute deviation (MAD). In robust statistical methods \footnote{https://en.wikipedia.org/wiki/Robust\_statistics}, MED is the robust measure of central tendency, while mean is not; MAD is robust measure of statistical dispersion, while standard deviation is not. Specifically, the $i$th input stock data is normalized to:
\begin{equation}
    \hat{\Mat{x}}[i] = |{\Mat{x}}[i] - \rm{MED}({\Mat{X}})| / \rm{MAD}({\Mat{X}}).
\end{equation}

\subsubsection{Dropping Extreme Label} When a stock hits limit up(limit down), stockholders are more reluctant to sell (buy) at this time, for they expect that the trend continues; as a result, it is difficult for other stockholders to buy (sell). Therefore, it is meaningless for the model to learn to ``buy when there is a limit up, and sell when there is a limit down''. To tackle this challenge, we propose to drop the extreme label to exclude the influence of extreme values. It is achieved in two ways: 1) we set a upper threshold and a lower threshold; 2) we drop the first and the last few percent labels.

\subsubsection{Software and Hardware} DiffsFormer is implemented with Python 3.8.16, Pytorch 1.11.0. All of the backbones are implemented with the open-sourced code in Qlib. We run the experiments on servers equipped with NVIDIA Tesla V100 GPU and 2.50GHz Intel Xeon Platinum 8163 CPU.

\begin{table}[t]
\setlength{\tabcolsep}{1.4mm}{
\caption{Weighted IC Comparison on CSI800 and CSI300.}
\label{tab:wic}
\begin{tabular}{c|cc|cc}
\toprule
  & \multicolumn{2}{c}{\textbf{CSI300}} & \multicolumn{2}{c}{\textbf{CSI800}}                                                                             \\ \cmidrule(l){2-5} 
 \textbf{Methods}                        & original & \multicolumn{1}{c|}{augmentation} & original    & augmentation \\ \midrule
\textbf{MLP}                         & 0.0326\tiny{±0.0023} &  \multicolumn{1}{c|}{\textbf{0.0332\tiny{±0.0021}}} & 0.0052\tiny{±0.0041}     & \textbf{0.0063\tiny{±0.0032}} \\
\textbf{LSTM}                & 0.0295\tiny{±0.0032} &  \multicolumn{1}{c|}{\textbf{0.0339\tiny{±0.0025}}} & \textbf{0.0075\tiny{±0.0055}}     & 0.0024\tiny{±0.0026} \\
\textbf{GRU}                        & 0.0324\tiny{±0.0012} &  \multicolumn{1}{c|}{\textbf{0.0383\tiny{±0.0011}}} & 0.0005\tiny{±0.0027}     & \textbf{0.0128\tiny{±0.0029}} \\

\textbf{SFM}                        & 0.0288\tiny{±0.0029} &  \multicolumn{1}{c|}{\textbf{0.0300\tiny{±0.0030}}} & \textbf{0.0028\tiny{±0.0032}}     & 0.0026\tiny{±0.0030} \\
\textbf{GAT}                        & \textbf{0.0354\tiny{±0.0006}} &  \multicolumn{1}{c|}{0.0324\tiny{±0.0004}} & \textbf{0.0083\tiny{±0.0010}}     & 0.0047\tiny{±0.0008} \\
\textbf{ALSTM}                        & 0.0260\tiny{±0.0038} &  \multicolumn{1}{c|}{\textbf{0.0312\tiny{±0.0033}}} & 0.0025\tiny{±0.0064}     & \textbf{0.0094\tiny{±0.0023}} \\
\textbf{HIST}                        &  0.0249\tiny{±0.0066}&  \multicolumn{1}{c|}{\textbf{0.0317\tiny{±0.0026}}} & -     & - \\
\textbf{Transformer}                        & 0.0387\tiny{±0.0038} &  \multicolumn{1}{c|}{\textbf{0.0433\tiny{±0.0048}}} & 0.0066\tiny{±0.0058}     & \textbf{0.0159\tiny{±0.0054}} \\

\bottomrule
\end{tabular}%
}
\vspace{-10pt}
\end{table}

\subsubsection{Model Parameters}
In Table \ref{tab:para}, we summarize the main hyper-parameters of DiffsFormer, as well as their search ranges. The optimal hyper-parameter are highlighted.

\subsection{Experimental Setup}
\subsubsection{Baselines} To verify the performance of the proposed framework in stock forecasting, we employ eight commonly used machine learning models as forecasting backbones:
\begin{itemize}
    \item \textbf{MLP}: a 2-layer multi-layer perceptron (MLP) with the number of units on each layer is 256.
    \item \textbf{LSTM} \cite{DBLP:journals/neco/HochreiterS97}: a Long Short-Term Memory network based stock price forecasting method.
    \item \textbf{GRU} \cite{DBLP:journals/corr/ChungGCB14}: a Gated Recurrent Unit (GRU) network based stock price forecasting method.
    \item \textbf{SFM} \cite{DBLP:conf/kdd/ZhangAQ17}: a State Frequency Memory (SFM) network that decomposes the hidden states of memory cells into multiple frequency components to model different latent trading patterns.
    \item \textbf{GAT} \cite{DBLP:conf/iclr/VelickovicCCRLB18}: a graph is constructed with stocks as nodes, and two stocks have a relation when they share the same stock concept. Then graph attention network (GAT) is utilized to aggregate the neighbor embedding attentively.
    \item \textbf{ALSTM} \cite{DBLP:conf/ijcai/FengC0DSC19}: an LSTM variant that incorporates temporal attentive aggregation layer to aggregate information from hidden embeddings in previous timestamps.
    \item \textbf{Transformer} \cite{DBLP:conf/nips/VaswaniSPUJGKP17}: a transformer \cite{DBLP:conf/nips/VaswaniSPUJGKP17} based stock price forecasting model.
    \item \textbf{HIST} \cite{DBLP:journals/corr/abs-2110-13716}: a graph-based framework that mines the concept-oriented shared information from predefined concepts and hidden concepts. The proposed framework simultaneously utilize the stock's shared information and individual information.
\end{itemize}

\subsubsection{Evaluation Metrics}
Since our task is stock forecasting, we utilize \textbf{Annualized Return Ratio (RR)} as the primary evaluation metric. In practice, we simulate the stock trade with ``top30drop30'' strategy: ``top30'' means that we keep the stocks with top30 predicted scores; and ``drop30'' means that each stock will be droped if its score falls out of top30, regardless of its previous performance. Besides, we also use two widely used metrics: \textbf{the Information Coefficient (IC)} \cite{DBLP:conf/kdd/LinZL021} and \textbf{Ranked Information Coefficient (Rank IC)} \cite{DBLP:conf/kdd/LiYZBQL19}. IC denotes the Pearson correlation between prediction and label, while Rank IC denotes the Spearman’s rank correlation between prediction and label. To eliminate the effect of random seed and other cause of performance fluctuation, we run the training and testing procedure 8 times for all of the methods and report the average value and the standard deviation. Since the training of DMs and the predictor is decoupled, we only run DM once for time efficiency.

\subsection{Performance Comparison}
To answer \textbf{RQ1}, we perform a completed comparison between the original and the augmented feature on the mentioned baselines, wherein the percentage of relative improvement on each metric is shown in Table \ref{tab:csi300} and \ref{tab:csi800}. Note that HIST requires the concept of stocks to build the graph, therefore we don't run it on CSI800 where the concepts are not available. Another notion is that the test time range is 2017-01-01 to 2020-12-31 in previous works \cite{DBLP:journals/corr/abs-2110-13716, DBLP:journals/corr/abs-2212-08656}, which is not consistent with \textbf{2020-04-01} to \textbf{2022-09-30} in our work. The reason is that we find factors and model performance can decays with age, and we aim to provide with an up-to-date performance of the models. As a result, the performance of backbones in our paper and that in previous works are not comparable. The main observation are as follows:
\begin{itemize}
    \item In general, the proposed framework DiffsFormer improve the performance of backbone models by $0.50\%\sim13.19\%$ and $4.01\%\sim70.84\%$ on return ratio on CSI 300 and CSI 800, respectively. This observation empirically verifies the necessity of data augmentation strategy in stock forecasting.
    \item Return ratio is the primary performance metric since the ultimate goal of stock forecasting is to achieve substantial profits. We observe that IC and RankIC are not positively associated to return ratio. We attribute the cause to our dataset CSI300 and CSI800 where going short is banned. Hence accurate modeling of tail stocks has little contribution to return ratio compared to that of top stocks. In alternative, we introduce to apply an exponentially decayed weight on IC to better characterize the correlation between the prediction and label on top stocks. The weighted result is reported in Table \ref{tab:wic}. From the table, we can observe that: 1) weighted IC for CSI800 is obviously lower than that for CSI300, which is consistent with return ratio performance in Table \ref{tab:csi300} and \ref{tab:csi800}. 2) The models' rankings in terms of weighted IC and return ratio are similar, especially on CSI800, suggesting weighted IC can be served as a metric to measure the potential of reaching a high return ratio. 3) DiffsFormer boosts the weighted IC for most of the methods on the CSI300 and improves the weighted IC for more than half of the methods on the CSI800, verifying its effectiveness of improving model performance.
    \item Due to the low SNR of data, we surprisingly observe that the performance of backbone methods with original features are counterintuitive (\eg ALSTM, as well as HIST which is one of the best methods in stock forecasting, are inferior to LSTM). Data augmentation to some extent gets the performance level back to normal. This observation suggests that data augmentation could alleviate the low SNR issue of financial data.
\end{itemize}

\begin{figure*}[t]
    \centering
    \begin{subfigure}[b]{0.33\textwidth}
    \centering
    \includegraphics[width=\textwidth]{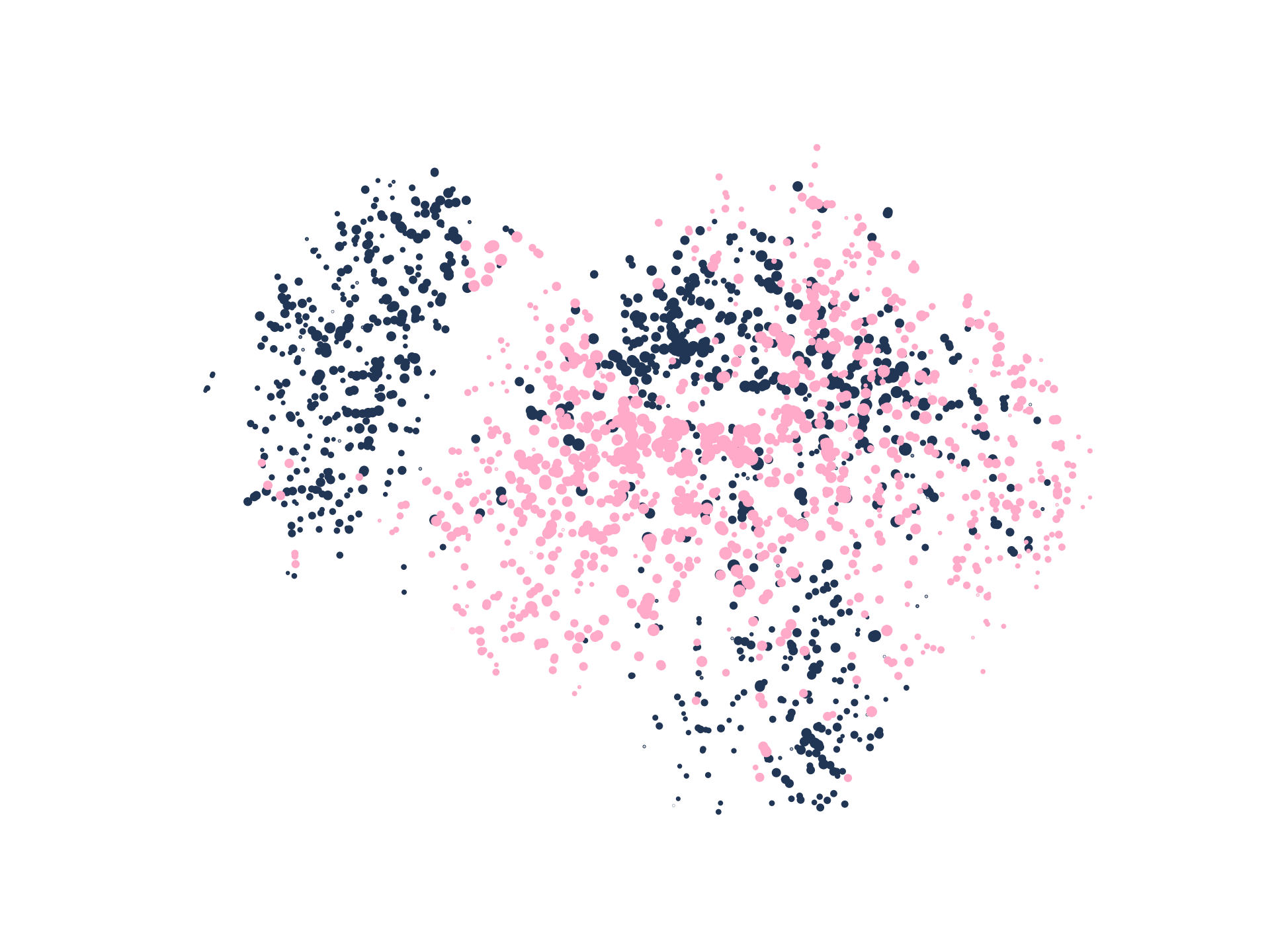}
    \caption{Direct Generation}
    \label{fig:dwog}
    \end{subfigure}%
    \begin{subfigure}[b]{0.33\textwidth}
        \centering
        \includegraphics[width=\textwidth]{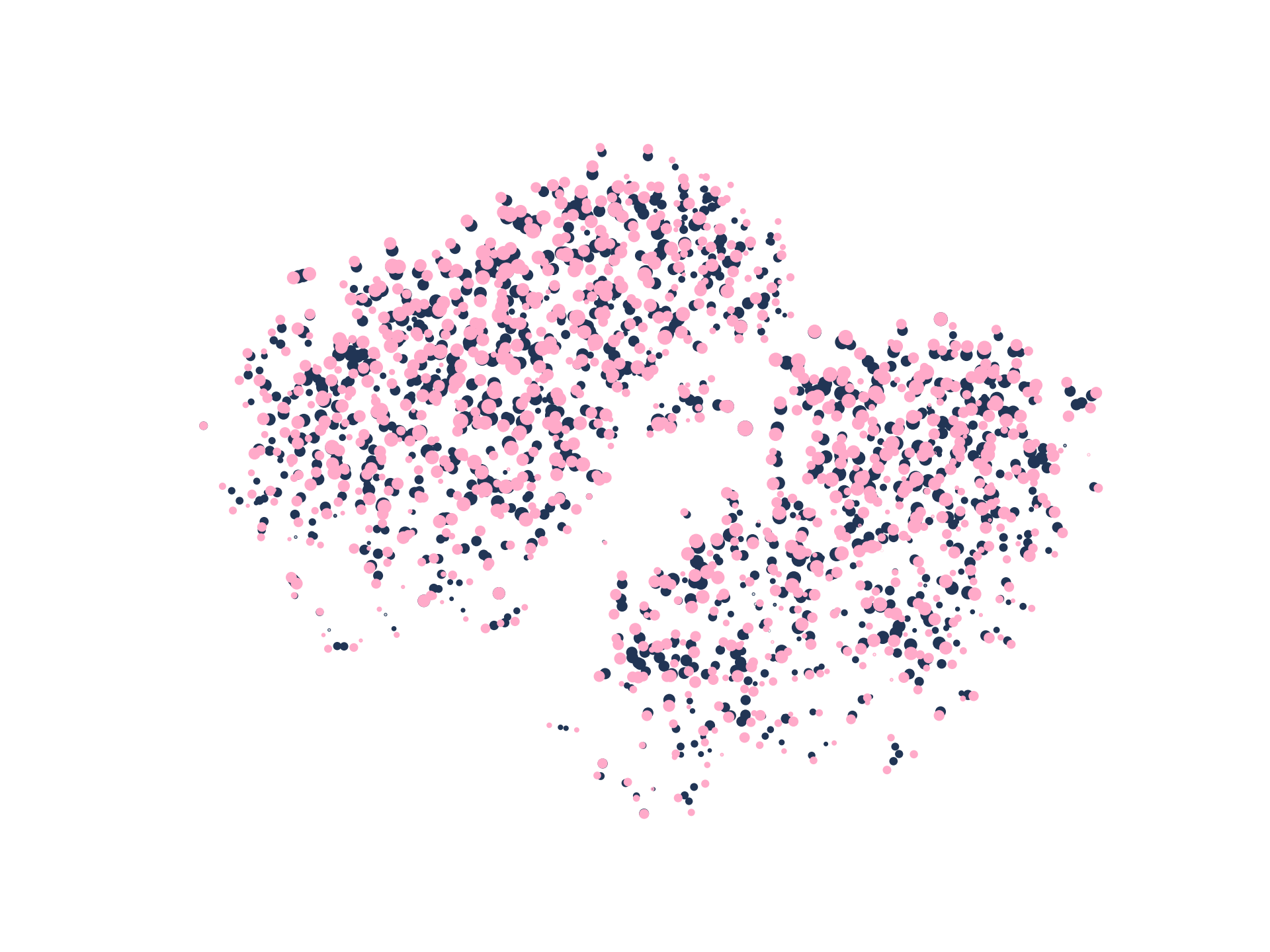}
        \caption{Random Noise Addition}
        \label{fig:rna}
    \end{subfigure}%
    \begin{subfigure}[b]{0.33\textwidth}
        \centering
        \includegraphics[width=\textwidth]{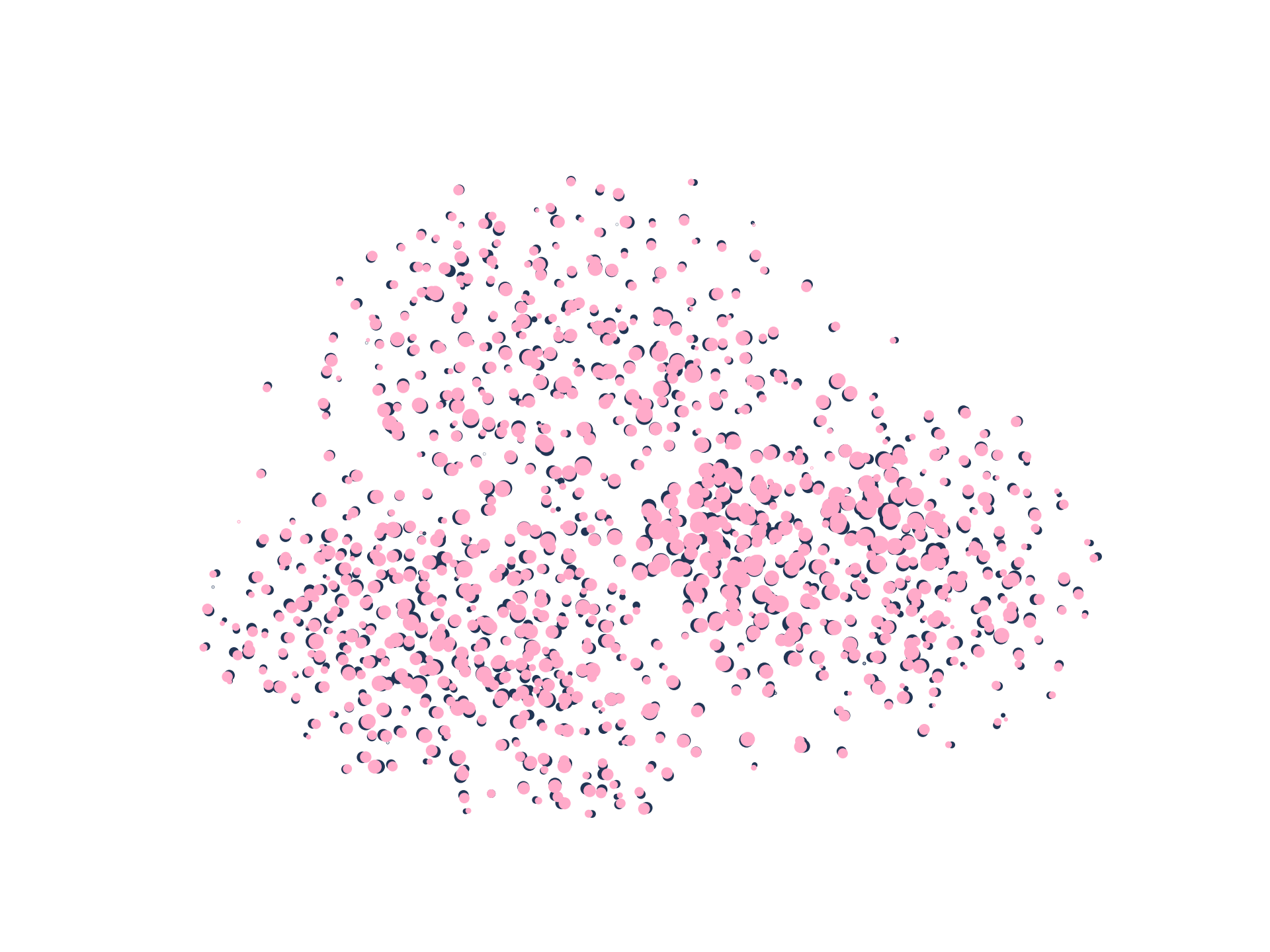}
        \caption{Editing}
        \label{fig:dwg}
    \end{subfigure}

    \caption{t-SNE plots of original features (pink) and augmented features (blue) elucidating the effect of editing, (b) denotes random gaussian addition. The lack of separation in the feature space suggests that the edited factors are meaningful.}
    \label{fig:tsne-feature-label}
    \vspace{-10pt}
\end{figure*}

\subsection{Discussion}
To answer \textbf{RQ2}, we elaborate on the quality of the edited data, and investigate the effectiveness of our proposed novel components.

\subsubsection{Data Fidelity and Diversity} 
\begin{figure}[t]
    \centering
    \begin{subfigure}[b]{0.5\columnwidth}
    \centering
    \includegraphics[width=\textwidth]{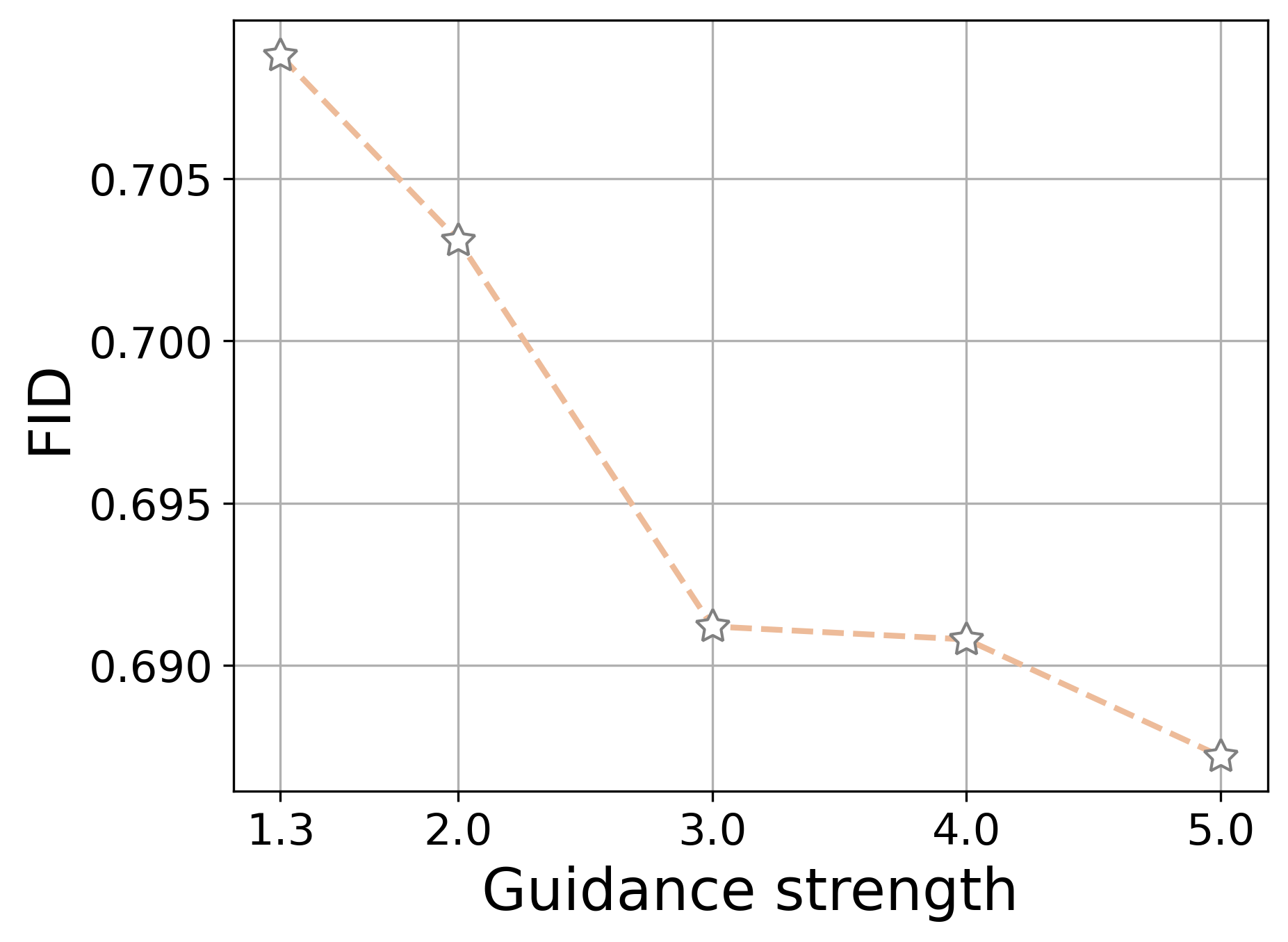}
    \caption{FID \wrt Strength}
    \label{fig:strength-fid}
    \end{subfigure}%
    \begin{subfigure}[b]{0.5\columnwidth}
        \centering
        \includegraphics[width=\textwidth]{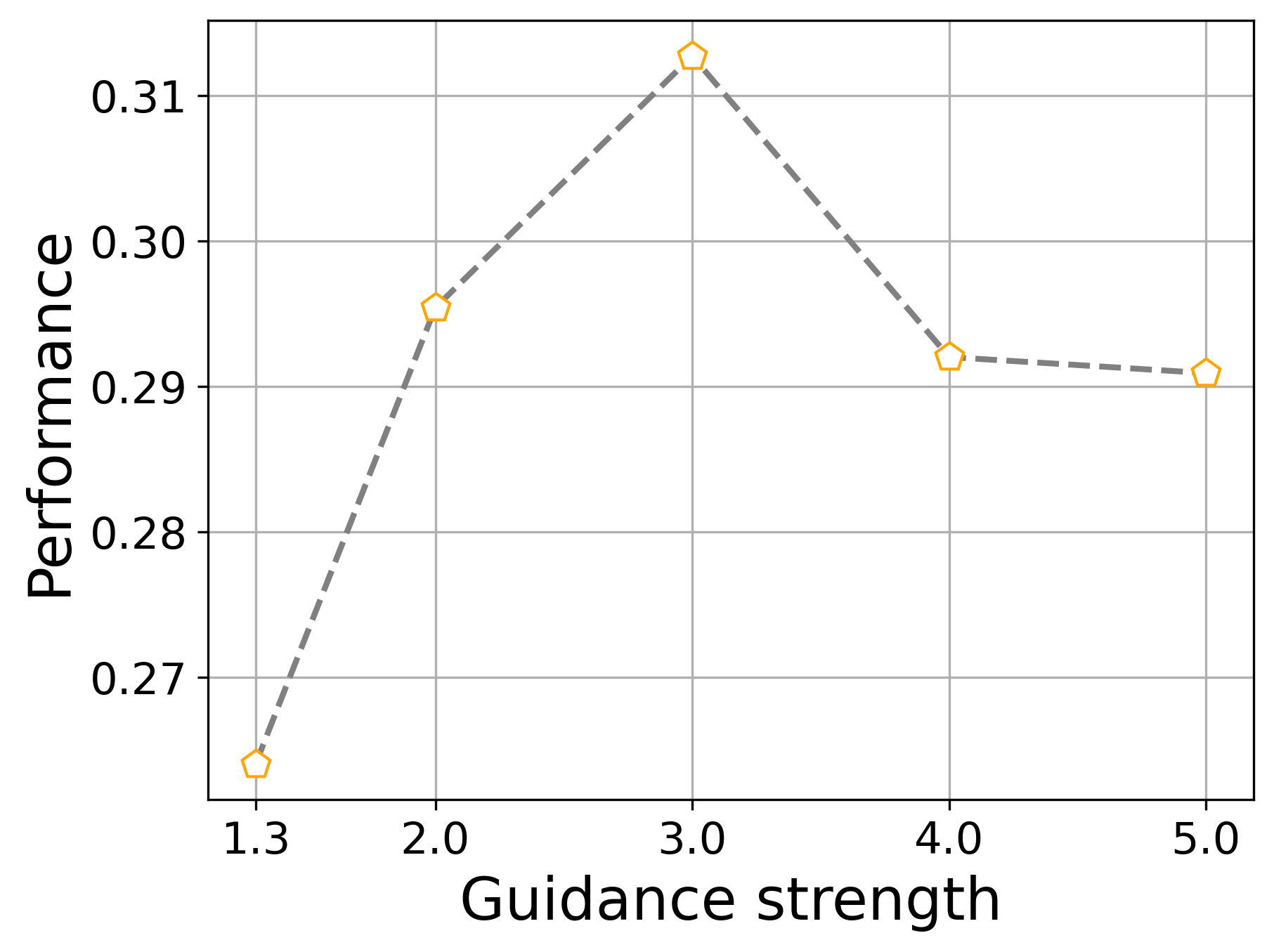}
        \caption{Performance \wrt Strength}
        \label{fig:strength-performance}
    \end{subfigure}
    \caption{Illustration of Data Fidelity and Diversity}
    \label{fig:fidelity}
    \vspace{-10pt}
\end{figure}

Our work adopts diffusion-based data augmentation module to synthesize data, which helps alleviate the serious data scarcity issue in stock forecasting. Particularly, before the start of training of the predictor in each epoch, we generate a new set of stock data. Therefore, the total amount of data utilized is \textbf{\textit{n$\times$}} the original where \textit{n} is equal to the total number of epoches. In other words, with DiffsFormer, the backbone can observe \textbf{\textit{n$\times$}} the data for once, instead of observing original data for $n$ times. To measure the fidelity and diversity of the data generated, two metrics are commonly adopted \cite{ho2020denoising, rombach2022high, peebles2022scalable}: Fréchet Inception Distance (FID) and Inception Score (IS). FID measures the distance between means and standards of two Gaussian distributions, given that two Gaussian distributions are close if their mean and standards are similar, hence a smaller FID indicates high fidelity; IS measures the KL Divergence between $p(y|x)$ and $p(y)$, given that $p(y|x)$ is sharp if data has high quality and $p(y)$ is flat if data has great diversity. However, $p(y|x)$ requires a well-trained classifier (\eg Inception Network), instead we use model performance to measure both the diversity and the fidelity of the data. As reported in Figure \ref{fig:strength-fid}, FID decreases as the guidance strength gets stronger, and DiffsFormer could achieve FID as low as 0.6872, suggesting a high fidelity of the generated data. This finding also verifies the effectiveness of editing mechanism, as DiffsFormer could achieve 0.7088 FID. In addition, from Figure \ref{fig:strength-performance}, we find that strong guidance strength may lead to performance drop. We attribute the reason for the phenomenon to the lack of diversity of the data.

\begin{table}[t]
\centering
\setlength{\tabcolsep}{3mm}{
\caption{The Effect of Editing Steps}
\label{tab:corruption}
\begin{tabular}{c|c|c|c|c}
\toprule
Editing Step      & 200     & 300   &  400    &  500  \\ \midrule
Performance & 0.284258 & 0.312679 & 0.293627 & 0.271165 \\ \midrule
FID & 0.411302 & 0.690823 & 1.138044 & 1.892653 \\ \bottomrule
\end{tabular}%
}
\end{table}

\subsubsection{Editing vs. Generating}
Due to the low SNR of the data, we train DM in the source domain that is much larger than target domain. During inference, we start from data point in the target domain, and corrupt them for a few steps before reversing to obtain a new data point in the target domain. In Figure \ref{fig:tsne-feature-label}, following recent work \cite{shipard2020DDN}, we visualize the relationship between the augmented features and the original stock features in blue and pink, respectively. We have two observations: 1) Comparing Figure \ref{fig:dwog} and Figure \ref{fig:dwg}, we find generated data are restricted to locate near the original data when we edit the existing sample from the target domain; while many points deviate the target domain distribution when we directly synthesize new data points. 2) Random gaussian noise addition can be treated as a special augmentation mechanism. We run several experiments with different level random gaussian noise addition and plot in Figure \ref{fig:rna} the t-SNE of feature distribution with the most accurate return ratio prediction. Our proposed method looks better than random noise addition. 

Since the target domain is a subset of the source domain, we distill new knowledge and information and enhance the data heterogeneity. Editing step $T'$ during inference can control the strength of knowledge distillation: a larger $T'$ makes generated data resemble more feature distribution from the source domain, while a smaller $T'$ makes edited data closer to original target domain data. To support this argument, we report the editing steps along with corresponding model performance and FID between the original and the edited data in Table \ref{tab:corruption}. We observe a trade-off between model performance and the editing step, which we attribute to the increased data diversity in the very early diffusion steps and the decreased data fidelity in the later steps.

\begin{table}[t]
\setlength{\tabcolsep}{3.4mm}{
\caption{Data augmentation and fine-tuning results on different target and source domains.}
\label{tab:transfer}
\begin{tabular}{@{}cc|c|c@{}}
\toprule
\multicolumn{2}{c|}{Mechanism}                               & Fine Tuning & Diffusion DA \\ \midrule
\multicolumn{1}{c|}{Target Domain}           & Source Domain &                 \multicolumn{2}{c}{Return Ratio}                         \\ \midrule
\multicolumn{1}{c|}{\multirow{2}{*}{CSI800}} & CSI800        &     0.1751\tiny{±0.0386}           &      0.1793\tiny{±0.0113}                        \\
\multicolumn{1}{c|}{}                        & CSIS          &                 0.1641\tiny{±0.0300}                   &   0.1903\tiny{±0.0382}                           \\ \midrule
\multicolumn{1}{c|}{\multirow{3}{*}{CSI300}} & CSI300        &     0.2789\tiny{±0.0376}           &        0.2861\tiny{±0.0547}                      \\
\multicolumn{1}{c|}{}                        & CSI800        &                 0.2773\tiny{±0.0181}                   &   0.2789\tiny{±0.0333}                           \\
\multicolumn{1}{c|}{}                        & CSIS          &                 0.2432\tiny{±0.0372}                   &   0.3127\tiny{±0.0113}              \\ \bottomrule
\end{tabular}%
}
\end{table}

\begin{table}[t]
\caption{The Stock Forecasting Performance with Different Conditionings}
\label{tab:conditions}
\begin{tabular}{@{}cc|cc@{}}
\toprule
\multicolumn{2}{c|}{}                                                          & Performance & FID \\ \midrule
\multicolumn{2}{c|}{Original}                                                  &  0.278941           &    - \\
\multicolumn{2}{c|}{Unconditioned}                                             &  0.291862           &  0.900919   \\
\multicolumn{2}{c|}{Predictor Guidance}                                        &            0.302396 &   0.900562  \\ \midrule
\multicolumn{1}{c|}{\multirow{3}{*}{Predictor-free Guidance}} & Label          & 0.297099 &  0.733522   \\
\multicolumn{1}{c|}{}                                         & Industry       &            0.300917 &  0.822593   \\
\multicolumn{1}{c|}{}                                         & Label+Industry &            \textbf{0.312679} &  \textbf{0.690823}   \\ \bottomrule
\end{tabular}
\vspace{-5pt}
\end{table}

\begin{figure*}[t]
    \centering
    \begin{subfigure}[b]{0.33\textwidth}
        \centering
        \includegraphics[width=\textwidth]{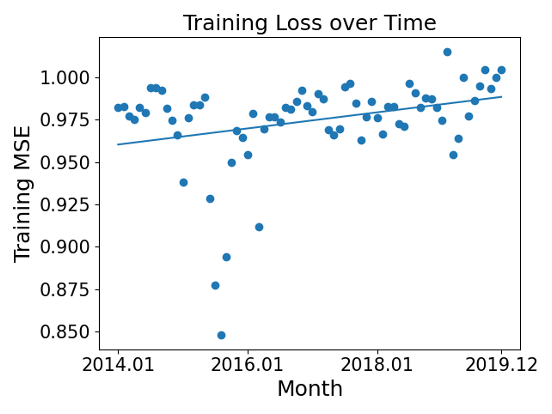}
        \caption{Original Training Loss \wrt Months}
        \label{fig:ori_loss}
    \end{subfigure}%
    \begin{subfigure}[b]{0.33\textwidth}
        \centering
        \includegraphics[width=\textwidth]{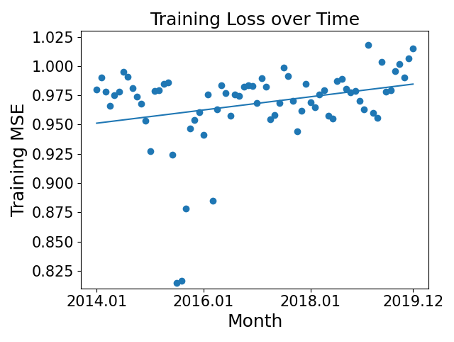}
        \caption{Uniform noise addition}
        \label{fig:uni_loss}
    \end{subfigure}%
    \begin{subfigure}[b]{0.33\textwidth}
        \centering
        \includegraphics[width=\textwidth]{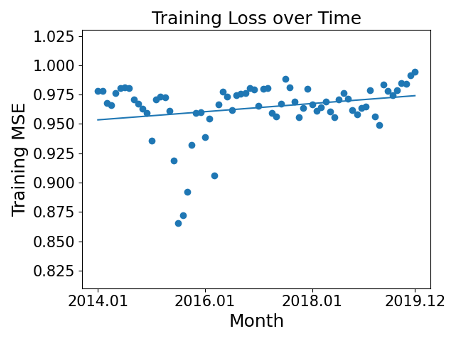}
        \caption{Loss-guided noise addition}
        \label{fig:guide_loss}
    \end{subfigure}
    \caption{The illustration of the impact of loss-guided diffusion.}
    \label{fig:loss-guidance}
    \vspace{-10pt}
\end{figure*}
\begin{figure}[t]
\centerline{\includegraphics[width=0.5\textwidth]{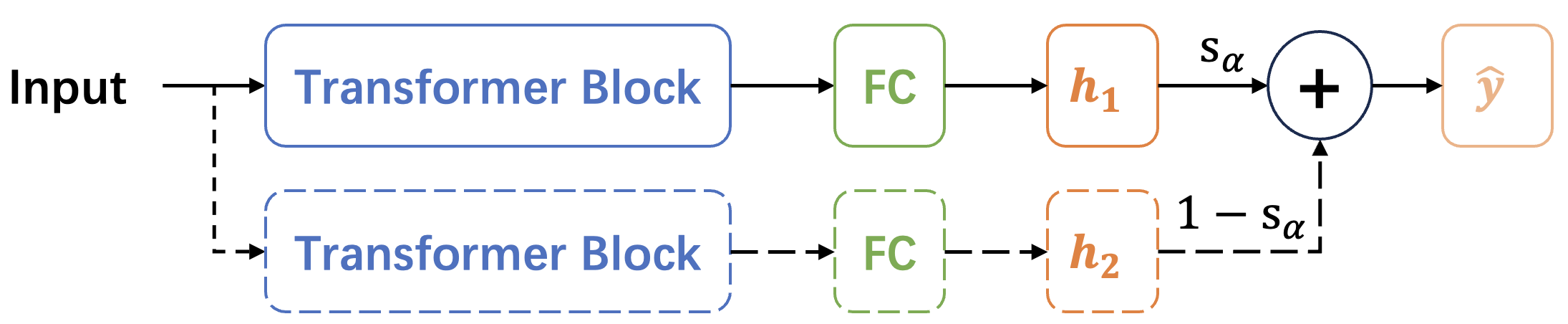}}
\caption{Illustration of Shake-shake with Transformer.}
\vspace{-10pt}
\label{fig:shake}
\end{figure}

\subsection{Effectiveness Analysis}
\subsubsection{Transfer Diffusion}
Recall that in \S \ref{sec:dda}, we design a novel inference process to distill new knowledge to generated data through transfer learning. To verify the real cause of performance improvement, we aim to exclude the interference of the new information. The result is shown in Table \ref{tab:transfer}, where the fine tuning column denotes the mechanism of training on source domain and testing on target domain, and diffusion DA stands for diffusion-based data augmentation with transfer learning. Our observation are three-folds: 1) Training in a larger source domain and testing in the target domain may degrade the model performance although new information is introduced, we suppose the reason is the distribution difference between the source domain and the target domain. 2) When the source domain and the target domain are the same, which means no new information is involved, the DM still can help raise the performance. 3) The transfer diffusion boosts the model performance by a large margin, demonstrating the effectiveness of the transfer learning mechanism.

\subsubsection{Conditional Diffusion} To adapt DM from generative task to regression task, we adopt predictor guidance and predictor-free guidance approaches. The stock forecasting performance with different conditionings are reported in Table \ref{tab:conditions}. We observe that DMs achieve lower FID and contribute to a better model performance with the help of conditionings.

\subsubsection{Comparison with Other Augmentation Algorithms}
\begin{figure}[t]
\centerline{\includegraphics[width=0.5\textwidth]{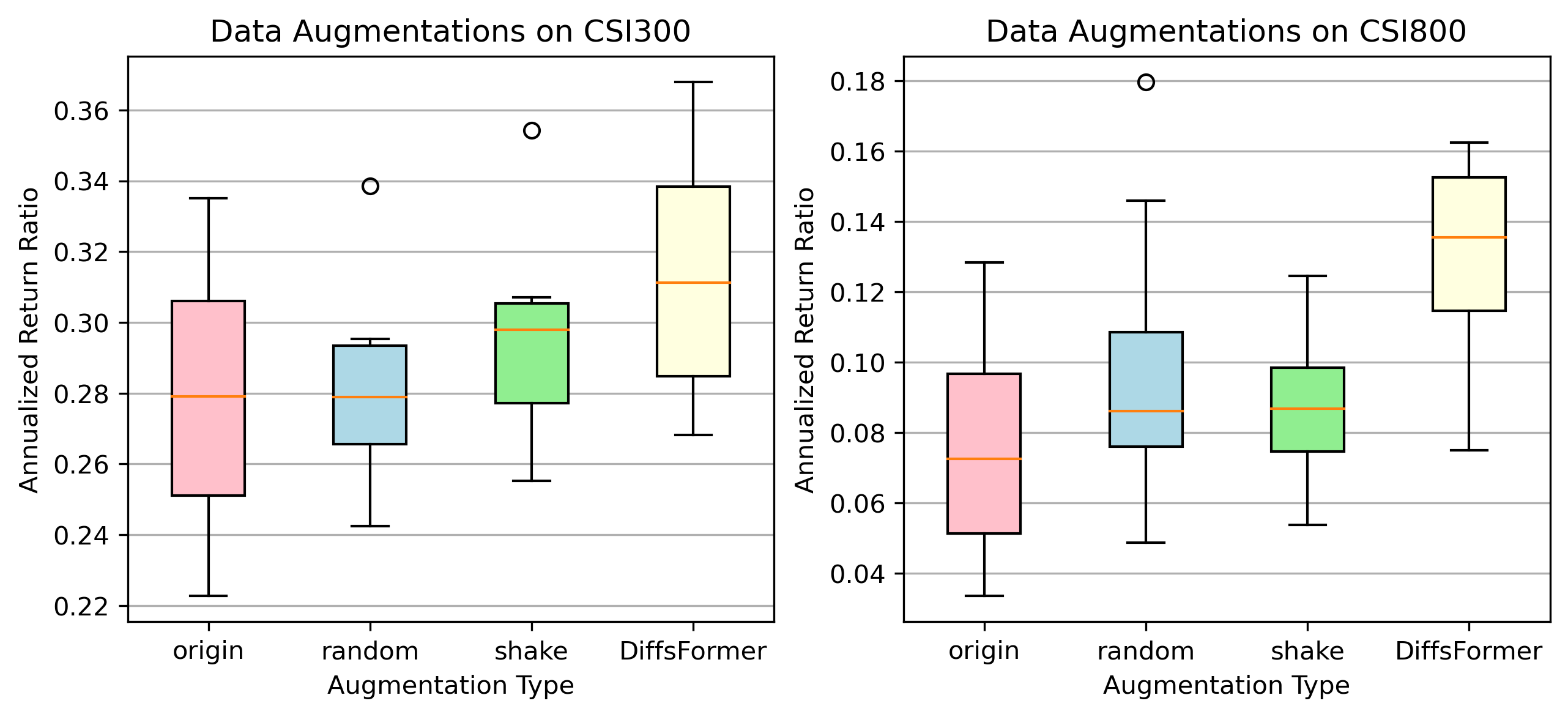}}
\caption{Comparison between different augmentation methods on CSI300 with Transformer and CSI800 with GRU.}
\vspace{-10pt}
\label{fig:box}
\end{figure}

In this work, we reveal that data augmentation plays a pivotal role in stock forecasting. And in this section, we aim to verify the DiffsFormer's superiority over other data augmentation mechanisms. The experimental results are reported in Figure \ref{fig:box}. The box from left to right are: 1) origin feature; 2) random gaussian noise addition; 3) shake-shake augmentation; 4) DiffsFormer. Shake-shake regularization \cite{DBLP:journals/corr/Gastaldi17} is first proposed to help alleviate the overfitting problem in deep learning. The core idea is to replace the standard forward layer with a stochastic affine combination of parallel branches; to make it clearer, a toy illustration example is shown in Figure \ref{fig:shake}. In each epoch, we sample $s_{\alpha}$ stochastically and \textit{keep} it instead of resampling $s_{\beta}$ for backward propagation. From this perspective, shake-keep could be treated as doubling the original data. From Figure \ref{fig:box}, we observe that: 1) Directly tweaking feature with noise addition can bring performance improvement, indicating that data collision is severe in dataset and the SNR of the data is quite low. 2) Shake-shake and DiffsFormer are two effective data augmentation mechanisms that outperform the random gaussian noise addition, and our proposed method DiffsFormer performs better than Shake-shake by a large margin. 3) Data augmentation can enhance the model stability, as the box of the augmentation is commonly shorter than that of the original. 4) The lower bound of the algorithm increases from the left to the right, suggesting that data augmentation improve the worst-case model performance.

\subsection{Improvements for DiffsFormer}
\subsubsection{Loss-guided Noise Addition}
\label{sec:loss-guidance}
\begin{figure}[t]
\centerline{\includegraphics[width=0.5\textwidth]{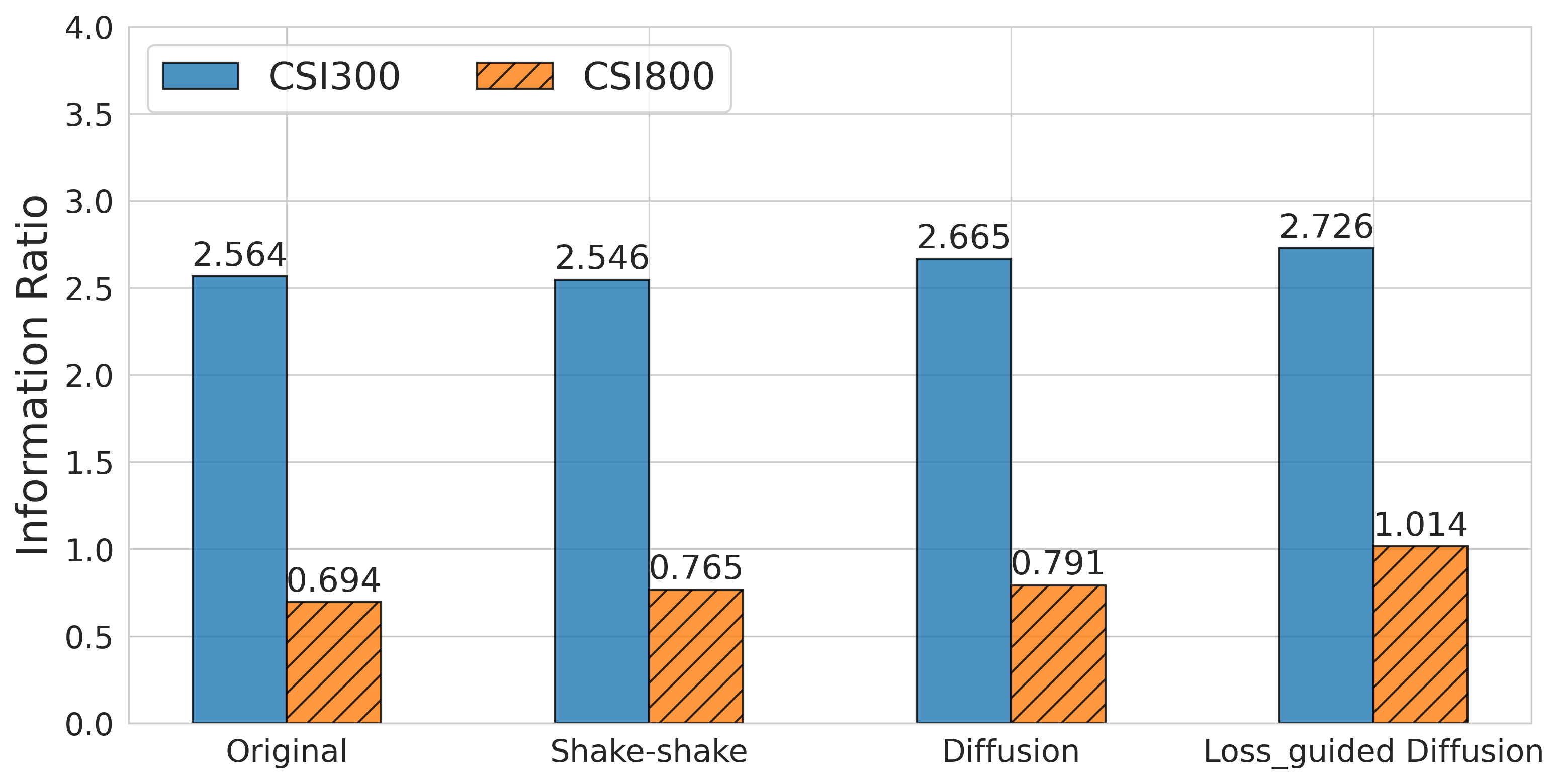}}
\caption{Information ratio by different training data.}
\vspace{-10pt}
\label{fig:loss-guided-bar}
\end{figure}

Besides the annualized return ratio, information ratio (IR) is another essential measure of the stock forecasting performance. IR is the ratio of the active return of the portfolio divided by the tracking error of its return, where tracking error identifies the level of consistency with which a portfolio "tracks" the performance of an index\footnote{https://en.wikipedia.org/wiki/Information\_ratio}. In other words, it measures the stability and generalization of the model. We discover that there exists some easy-fitted points in the dataset, and we assume that alleviating the overfitting issue of these extreme data points can decrease volatility and lead to a higher information ratio. To answer \textbf{RQ3}, we plot the training loss over time in Figure \ref{fig:loss-guidance}. The loss for stock forecasting is quite low during the stock crash that occurred between June 2015 to June 2016, we suspect that the cause is the increased proportion of retail investment whose action pattern is simpler. A model that fits too much to the data around 2015 will fail in a recent time as the market pattern becomes more complex; however, it is sub-optimal to remove these data which exaggerates the data scarcity. Thus, we propose a novel strategy called loss-guided noise addition. Specifically, we leverage training loss as a proxy and add stronger noise to data points with lower training loss. As shown in Figure \ref{fig:guide_loss}, the loss-guided diffusion makes the training losses flatter compared to uniform noise-addition, alleviating the overfitting issue and achieving higher information ratio as verified in Figure \ref{fig:loss-guided-bar}, where IR on CSI300 and CSI800 with different data are reported. We observe that (1) data augmentation can increase the IR of the model; (2) DM outperforms shake-shake; (3) loss-guided diffusion can further increase IR and decrease the volatility.

\subsubsection{Time Efficiency Improvement}
\begin{figure}[t]
\centerline{\includegraphics[width=0.5\textwidth]{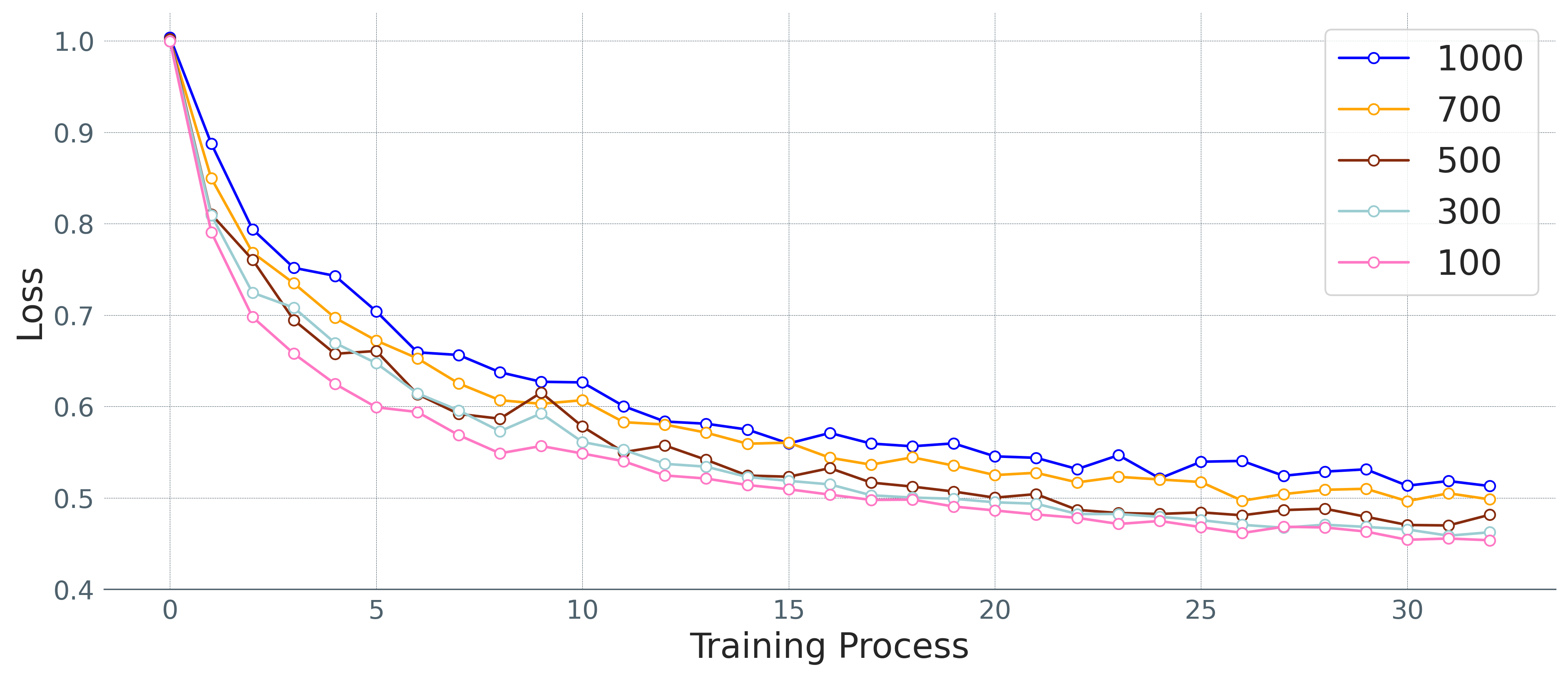}}
\caption{Loss curves with different sampling steps.}
\vspace{-10pt}
\label{fig:smoothed-loss}
\end{figure}

From previous analysis in \S \ref{sec:dda}, it is obvious to see that there is no need to optimize $\epsilon_{\theta}(\Mat{x}_{t}, t)$ for $t > T'$ under our transfer learning framework. Since DMs are time-consuming, we develop a trick to speed up the training of the framework. Concretely, we initialize $\alpha$ and $\beta$ with total diffusion steps $T$ to ensure correctness; however, we sample \textit{training step} t from $Uniform\{1, 2, \cdots, T'\}$ instead of $Uniform\{1, 2, \cdots, T\}$. The loss curves with maximum sampling steps within the set \{100, 300, 500, 700, 1000\} are elucidated in Figure \ref{fig:smoothed-loss}. Note that the figure represents the average loss within sampling step 100 instead of diffusion step 1000. We discover that with the decrease of sampling steps, DMs embrace with a more sharp loss curve, which means they can converge faster.

\subsubsection{Data Collision}
In modern stock market, several models which utilize the same copy of data coexist. If they provide with the same recommendation, the influx of more money into the market will make some factors ineffective, and models relying on these factors get failed \cite{jacobs2020anomalies}. In Table \ref{tab:collision}, we report the performance of model with three types of input: the original data, the augmented data, and the union of these two datasets. We have several observations: 1) Data augmentation is effective since it boosts the model performance; 2) Utilizing both the original data and the augmented data is inferior to using augmented data only. Given the assumption that original data has better data quality than the augmented data, and our test data is real-world, we guess the reason is data collision, which our proposed model can address. Furthermore, in Figure \ref{fig:box}, we surprisingly observe that directly tweaking feature (\eg applying noise) can bring performance improvement, further indicating that data collision is severe in dataset.

\begin{table}[t]
\setlength{\tabcolsep}{2.2mm}{
\caption{Performance with different inputs.}
\label{tab:collision}
\begin{tabular}{c|c|c|c}
\toprule
            & Original only   &  Union of the two    &  Augmented only     \\ \midrule
Performance & 0.278941 & 0.298965 & 0.312679 \\ \bottomrule
\end{tabular}%
}
\vspace{-5pt}
\end{table}

%% file: chapters/6_conclusion.tex
\section{Conclusion and Future Work}
\label{sec:conclusion}
In this work, we reveal that one of the most challenges that stock forecasting task faces is data scarcity. To address the issue, we propose DiffsFormer, a novel conditional diffusion Transformer framework for stock forecasting, which focuses on augmenting time-series stock data with help of label and industry information. To distill new knowledge and information, we incorporate transfer learning in DM by training it in a larger source domain and synthesizing with data in target domain. Furthermore, we develop several novel mechanisms to boost the overall model performance, decrease the volatility, increase the time efficiency.

This work takes a first step to do data augmentation in stock forecasting with diffusion. There are some future work directions and limitations can provide valuable insights for further research. We discover that conditionings such as industry sector can improve the performance, hence it is possible to enhance the performance of target stocks by editing factors into a specific industry or generating stocks with specific market value. Additionally, this work highlights the issues of data collision and homogeneity in stock forecasting, further research could involve developing techniques to identify and handle data collisions, as well as strategies to introduce diversity and heterogeneity in the training data explicitly. 